\documentclass[fleqn,usenatbib]{mnras}

\usepackage{newtxtext,newtxmath}
\usepackage[T1]{fontenc}

\DeclareRobustCommand{\VAN}[3]{#2}
\let\VANthebibliography\thebibliography
\def\thebibliography{\DeclareRobustCommand{\VAN}[3]{##3}\VANthebibliography}

\usepackage{graphicx}	% Including figure files
\usepackage{amsmath}	% Advanced maths commands
\usepackage[section]{placeins}
\usepackage[table, svgnames, dvipsnames]{xcolor}
\usepackage{adjustbox}
\usepackage{rotating,booktabs}
\newbox\grsign \setbox\grsign=\hbox{$>$} \newdimen\grdimen \grdimen=\ht\grsign
\newbox\simlessbox \newbox\simgreatbox
\setbox\simgreatbox=\hbox{\raise.5ex\hbox{$>$}\llap
     {\lower.5ex\hbox{$\sim$}}}\ht1=\grdimen\dp1=0pt
\setbox\simlessbox=\hbox{\raise.5ex\hbox{$<$}\llap
     {\lower.5ex\hbox{$\sim$}}}\ht2=\grdimen\dp2=0pt

\def\simless{\mathrel{\copy\simlessbox}}
\newbox\simppropto
\setbox\simppropto=\hbox{\raise.5ex\hbox{$\sim$}\llap
     {\lower.5ex\hbox{$\propto$}}}\ht2=\grdimen\dp2=0pt

\usepackage{xcolor}
\title[Terzan~5 according to APOGEE]{Is Terzan~5 the remnant of a building block of the Galactic bulge? Evidence from APOGEE}

\author[D. J. Taylor et al.]{Dominic J. Taylor,\textsuperscript{1}\thanks{E-mail: \href{mailto:dom.taylor111@gmail.com}{dom.taylor111@gmail.com}}
Andrew C. Mason,\textsuperscript{1}
Ricardo P. Schiavon,\textsuperscript{1}
Danny Horta,\textsuperscript{1}
David M. Nataf,\textsuperscript{2}
\newauthor Doug Geisler,\textsuperscript{3,4,5}
Shobhit Kisku,\textsuperscript{1}
Si$\hat{\textrm{a}}$n G. Phillips,\textsuperscript{1}
Roger E. Cohen,\textsuperscript{6}
Jos\'e G. Fern\'andez-Trincado,\textsuperscript{7}
\newauthor Timothy C. Beers,\textsuperscript{8}
Dmitry Bizyaev,\textsuperscript{9,10}
Domingo An\'ibal Garc\'ia-Hern\'andez ,\textsuperscript{11}
Richard R. Lane,\textsuperscript{12}
\newauthor Pen\'elope Longa-Pe$\Tilde{\textrm{n}}$a,\textsuperscript{13}
Dante Minniti,\textsuperscript{14,15}
Cesar Mu$\Tilde{\textrm{n}}$oz,\textsuperscript{4,16}
Kaike Pan,\textsuperscript{9}
and Sandro Villanova\textsuperscript{6}\\
\textsuperscript{1} Astrophysics Research Institute, Liverpool John Moores University, 146 Brownlow Hill, Liverpool L3 5RF, UK\\
\textsuperscript{2} Center for Astrophysical Sciences and Department of Physics and Astronomy, The Johns Hopkins University, Baltimore, MD 21218, USA\\
\textsuperscript{3} Departamento de Astronom\'ia, Universidad de Concepci\'on, Casilla 160-C, Concepci\'on, Chile\\
\textsuperscript{4} Instituto de Investigaci\'on Multidisciplinario en Ciencia y Tecnolog\'ia, Universidad de La Serena, Benavente 980, La Serena, Chile\\
\textsuperscript{5} Departamento de Astronom\'ia, Universidad de La Serena, Avenida Juan Cisternas 1200, La Serena, Chile\\
\textsuperscript{6} Space Telescope Science Institute, 3700 San Martin Drive, Baltimore, MD 21218, USA\\
\textsuperscript{7} Instituto de Astronom\'ia, Universidad Cat\'olica del Norte, Av. Angamos 0610, Antofagasta, Chile\\
\textsuperscript{8} Department of Physics and JINA Center for the Evolution of the Elements, University of Notre Dame, Notre Dame, IN 46556, USA\\
\textsuperscript{9} Apache Point Observatory and New Mexico State University, P.O. Box 59, Sunspot, NM, 88349-0059, USA\\
\textsuperscript{10} Sternberg Astronomical Institute, Moscow State University, Moscow, Russia\\
\textsuperscript{11} Instituto de Astrof\'isica de Canarias, 38205 La Laguna, Tenerife, Spain\\
\textsuperscript{12} Centro de Investigación en Astronomía, Universidad Bernardo O'Higgins, Avenida Viel 1497, Santiago, Chile\\
\textsuperscript{13} Centro de Astronom\'ia, Universidad de Antofagasta, Avenida Angamos 601, Antofagasta 1270300, Chile\\
\textsuperscript{14} Departamento de Ciencias F\'isicas, Universidad Andres Bello, Av. Republica 220, Santiago, Chile\\
\textsuperscript{15} Vatican Observatory, V00120 Vatican City State, Italy\\
\textsuperscript{16} Universidade de S$\Tilde{\textrm{a}}$o Paulo, IAG, Rua do Mat$\Tilde{\textrm{a}}$o 1226, Cidade Universit\'aria, S$\Tilde{\textrm{a}}$o Paulo 05508-900, Brazil
}

\date{Accepted XXX. Received YYY; in original form ZZZ}

\pubyear{2021}

\begin{document}
\label{firstpage}
\pagerange{\pageref{firstpage}--\pageref{lastpage}}
\maketitle

\begin{abstract}
It has been proposed that the globular cluster-like system Terzan~5 is the surviving remnant of a primordial building block of the Milky Way bulge, mainly due to the age/metallicity spread and the distribution of its stars in the $\alpha$-Fe plane.  We employ Sloan Digital Sky Survey (SDSS-IV) data from the Apache Point Observatory Galactic Evolution Experiment (APOGEE~2) to test this hypothesis.  Adopting a random sampling technique, we contrast the abundances of 10 elements in Terzan~5 stars with those of their bulge field counterparts with comparable atmospheric parameters, finding that they differ at statistically significant levels.  Abundances between the two groups differ by more than 1$\sigma$ in Ca, Mn, C, O, and Al, and more than 2$\sigma$ in Si and Mg.  Terzan~5 stars have lower [$\alpha/Fe$] and higher [Mn/Fe] than their bulge counterparts.  Given those differences, we conclude that Terzan~5 is not the remnant of a {\it major} building block of the bulge.  We also estimate the stellar mass of the Terzan~5 progenitor based on predictions by the Evolution and Assembly of GaLaxies and their Environments (EAGLE) suite of cosmological numerical simulations, concluding that it may have been as low as $\sim3\times10^8$~M$_\odot$ so that it was likely unable to significantly influence the mean chemistry of the bulge/inner disk, which is significantly more massive ($\sim10^{10}$~M$_\odot$).  We briefly discuss existing scenarios for the nature of Terzan~5 and propose an observational test that may help elucidate its origin.

%{\cyan{} [[We have to work on the conclusions part of the abstract once we are done with the text]]}
%We thus conclude that the APOGEE data argues against Terzan 5 being a remnant of a building block of the Galactic bulge.
\end{abstract}


%%%%%%%%%%%%%%%%%%%%%%%%%%%%%%%%%%%%%%%%%%%%%%%%%%

%%%%%%%%%%%%%%%%% BODY OF PAPER %%%%%%%%%%%%%%%%%%

\section{Introduction}
\label{sec:intro}

In the Lambda Cold Dark Matter ($\Lambda$-CDM) scenario, galaxy assembly takes place largely through hierarchical merging.
%the accretion of lower mass systems to form the massive galaxies we observe at present day.
Galaxy formation theory can in principle be constrained from observations of the stellar populations within the Milky Way (MW), the galaxy we can observe in greatest detail. Evidence for accretion in the MW has been seen in the past, including the Sagittarius dwarf spheroidal (Sgr dSph) identified by \citet{Ibata_1994} and the more recently discovered Gaia-Sausage-Enceladus \citep[GSE;][]{Belokurov_2018,Haywood_2018,Helmi_2018,Mackereth_2019}. In the Galactic bulge, in particular, \citet{Horta_2021} found evidence for the presence of the remnants of the early accretion event of a massive satellite they named Heracles.

The inner few kiloparsecs of the MW, a region that we refer to, by convention, as {\it the bulge}, concentrates an important fraction of the Galaxy's total stellar mass, holding important clues to its early formation. Yet its precise evolutionary history remains elusive.  The stellar population content of the bulge is complex, as it hosts populations from different Galactic components and with distinct chemodynamic properties \citep[e.g.,][]{Minniti_1995,Nataf_2017,Barbuy_2018}. The situation is exacerbated by difficult observational access due to crowding and severe extinction \citep[e.g.,][]{Minniti_2010}.

Globular clusters (GCs) are the oldest surviving stellar systems in the Galaxy, and as such they are considered tracers of the early formation history of the MW.  The bulge population of GCs is particularly interesting, including systems with a wide range of properties, \citep[e.g.,][]{Barbuy_1998,Schiavon_2017b,Geisler_2021}, with a few cases hosting stellar populations with a range of ages and metallicities \citep{Ferraro_2021}.  One of these latter GCs is Terzan~5, which is the focus of our study.

%intriguing cases of which Terzan 5 (Terzan~5) is a point in case. 
%GCs are massive tightly bound agglomerations of thousands to millions of stars which are among the oldest systems in the universe. Therefore, as one would expect, these stellar systems hold valuable information about the formation and evolution of the MW which allows us to understand how galaxies form and evolve generally throughout the universe. 

%Star cluster Terzan~5, residing within the bulge of the MW, was first accepted as a GC until \citet{Ferraro_2016} concluded that it cannot be genuine, as it shares only a few properties with GCs, and therefore should be regarded as a GC-like system.

A growing number of Galactic GCs have been found to exhibit star-to-star variations in metallicity, such as $\omega$ Centauri \citep[exhibiting similarly large variation;][]{Lee_1999,Pancino_2000,Ferraro_2004,Ferraro_2006,Bellini_2009,Bellini_2010,Bellini_2013,Villanova_2014}, M54 \citep{Carretta_2010}, M22 \citep{Marino_2009, Marino_2011,Marino_2012}, M62 \citep{Yong_2014}, NGC 6273 \citep{Johnson_2017,Pfeffer_2021}, and recently Liller 1 \citep{Ferraro_2021}. The origin of the complex chemistry of these systems is still not entirely understood.

Photometric and spectroscopic studies by \citet{Ferraro_2009} and \citet{Origlia_2010,Origlia_2013} identified the presence of a multi-peak metallicity distribution in Terzan~5. More recently, \cite{Ferraro_2016} used HST-based proper motions in order to bring the stellar population content of Terzan~5 into sharp relief, revealing the existence of two stellar populations widely separated in age.  \cite{Ferraro_2016} suggest that the mass of the Terzan~5 progenitor could have been as high as ${\rm 10^8-10^9~ M_\odot}$. \citet{Schiavon_2017b} showed that the multiple population (MP) phenomenon \citep[e.g.,][]{Renzini_2015,BastianLardo_2018}, characterised by the presence of light element abundance anti-correlations, is present in Terzan~5, indicating that some of its populations have chemistry similar to that of standard GCs.

Another intriguing property of Terzan~5 concerns the abundance patterns of its members.  \cite{Ferraro_2016} showed that the distribution of Terzan~5 stars in the $\alpha$-Fe plane tracks relatively closely that of the bulge field.  The distribution of a system's stellar populations in this plane is a useful diagnostic of its star formation history \citep{Greggio_1983,Mannucci_2005,Maoz_2011}. \cite{Ferraro_2016} showed that the change in the slope -- colloquially termed the "knee" -- of the [$\alpha$/Fe] vs. [Fe/H] relation occurs at a similar metallicity in the two populations. Such a correlation indicates a similarity in the chemical evolution of the systems, where a decline in the [$\alpha$/Fe] abundance ratio from the $\alpha$-enhanced, SN II-enriched, plateau has historically been attributed to the onset of Type Ia SNe \citep[][but see Mason et al. 2022, in prep.]{Tinsley_1979, Matteucci_1986}. It has been claimed by various authors that the metallicity at which the knee occurs is related to the system's stellar mass \citep[e.g.,][]{Tolstoy_2009} and the efficiency of star formation a galaxy achieved prior to the onset of pollution of the ISM by large amounts of Fe from SNe Ia. According to this scenario, a system that both forms stars which enrich the ISM in the $\alpha$-elements through core-collapse (CC) supernovae efficiently $\textit{and retains}$ these metals produces a distribution in the $\alpha$-Fe plane that is characterised by a high metallicity of the knee \citep[e.g.,][Mason et al. 2022, in prep.]{McWilliam1997}. This results in a correlation between metallicity of the knee and galaxy mass, as early star-formation rates of more massive galaxies are more likely to be higher since their potential wells are deeper. The similarity between Terzan~5 and the bulge field in this plane has thus been suggested to be indicative of a high mass for the progenitor of Terzan~5, which would in turn suggest that this system was an important contributor to the stellar mass content of the Galactic bulge \cite{Ferraro_2016}.

The above evidence led to the suggestion that Terzan~5 could be the fossil remnant of a primordial building block of the bulge of the MW. Galaxy bulge formation has been suggested to occur through rapid assembly at early epochs, followed by the evolution of a central disc/bar and its interactions on a longer timescale with substructures formed {\it in situ} \citep{Kormendy_2004,Immeli_2004,Shen_2010}. The role of a system such as Terzan~5 in this picture has yet to be determined.

The hypothesis that the progenitor of Terzan~5 has contributed significantly to the stellar mass budget of the Galactic bulge can be tested through chemical tagging based on a large number of precise elemental abundances for statistically significant samples from both systems.  This is the task we set out to perform. We report evidence, based on SDSS-IV/APOGEE-2 DR17 spectroscopy, that the detailed chemical composition of Terzan~5 stars differs from that of the bulge field populations in a statistically significant way.  In addition, we examine the prediction by the EAGLE simulations for the dependence of knee metallicity on stellar mass to argue that the progenitor of Terzan~5 was likely not a major contributor to the stellar content of the Galactic bulge.\\ 

\indent The structure of the paper is as follows. The samples of stars associated with Terzan~5 and the bulge are presented in Section~\ref{sec:datasample}. The analysis and results of the chemical abundance patterns of Terzan~5 and bulge field stars are presented in Section \ref{sec:analysisresults}. Those results are discussed in light of  existing scenarios for the origin of Terzan~5 in Section~\ref{sec:nature}. Our conclusions are summarised in Section~\ref{sec:conclusions}.

% \indent {\cyan [[This paragraph is quite informative, but it deals with the issue of whether Terzan 5 is accreted, or not.  It's not central to our quest, since we cannot place a constraint on that issue.  Maybe we should move it somewhere else?  Leave it here for the time being]] Further to work by \citet{McKenzie_2018} which used numerical simulations to try to explain the existence of the super-solar population within Terzan~5, they present hypotheses for it being the result of a Galactic GC (or member of a dwarf galaxy) interacting with a giant molecular cloud (GMC) 5 Gyr ago (and the remnant accreted) or the remnant of a GC-GC interaction to explain its $2^\textrm{nd}$ generation population largely concentrated around the cluster centre. Interestingly, \citet{Schiavon_2017b} concluded that Terzan~5 has abundance spreads and correlations similar to $\omega$ Cen suggesting that they occupy a region of parameter space between GCs and dwarf spheroidal galaxies.}

% \FloatBarrier
\section{Data and Sample}
\label{sec:datasample}

\begin{table*}
\centering
\caption{Summary of the parameters used to select candidate Terzan~5 stars, including cluster mean RA $\alpha_\textrm{Ter 5}$ and Dec $\delta_\textrm{Ter 5}$ in degrees, Jacoby radius $r_J$ in arcmins, mean heliocentric cluster radial velocity RV$_\textrm{Ter 5}$ and dispersion $\sigma_\textrm{RV}$ in km/s, RA proper motion $\mu_{\alpha}\cos(\delta)$ and Dec proper motion $\mu_{\delta}$ in mas/yr, mean proper motion dispersion $\sigma_\textrm{PM}$ in mas/yr, and Galactocentric cluster distance $R_\textrm{GC}$ in kpc.}
    \begin{tabular}{@{}ccccccccc@{}}
    \hline\hline
    $\alpha_\textrm{Ter 5}^\circ$ & $\delta_\textrm{Ter 5}^\circ$ & $r_J^{\prime}$ & RV$_\textrm{Ter 5}$ & $\sigma_\textrm{RV}$ & $\mu_{\alpha}\cos(\delta)$ & $\mu_{\delta}$ & $\sigma_\textrm{PM}$ & $R_\textrm{GC}$ \\
    \hline%
%    $17^\textrm{h}48^\textrm{m}4.80^\textrm{s}$  & $-24^\circ46^{\prime}45^{\prime\prime}$  & 25.37 & -82.3 & 19 & 1.65 \\
    267.0202  & $-$24.77906  & 25.37 & $-$82.57 & 15.5 & $-$1.9955 & $-$5.243 & 0.54 & 1.65 \\
    \hline\hline
    \end{tabular}
\label{tab1}
\end{table*}

This paper utilizes data from the 17$^\textrm{th}$ Data Release (DR17) of the Sloan Digital Sky Survey \citep[SDSS-IV;][]{Blanton_2017} Apache Point Observatory Galactic Evolution Experiment \citep[APOGEE~2;][]{Majewski_2017}. APOGEE~2 performs a detailed characterisation of the stellar populations of the Milky Way and its satellite companions using twin multi-fiber spectrographs \citep{Wilson_2019} attached to the 2.5-m Sloan Foundation Telescope at the Apache Point Observatory \citep[APO;][]{Gunn_2006} in New Mexico and the 2.5-m du Pont Telescope \citep{Bowen_1973} at the Las Campanas Observatory (LCO) in Chile. The high resolution  (R$\sim$23,000) spectra are collected in the near-infrared (NIR) H~band, yielding highly precise radial velocities and chemical compositions for over hundreds of thousands of stars across both hemispheres. The focus on the NIR is essential to investigate stars located in the Galactic disc and bulge due to the high extinction caused by intervening dust. We use atmospheric parameters, elemental abundances, and quality flags for stars from APOGEE~2, based on the automatic analysis of its spectra performed by the APOGEE Stellar Parameter and Chemical Abundances Pipeline \citep[ASPCAP;][]{Perez_2016,Holtzman_2015,Holtzman2018,Jonsson2020}\footnote{The analysis in this paper is based on the {\tt synspec-rev1} version of the catalogue.}, and stellar distances provided by an \texttt{astroNN} neural network trained on stars with APOGEE spectra and {\it Gaia} EDR3 \citep{Gaia_2016,Gaia_2018b} parallax measurements \citep{LeungBovy_2019a,LeungBovy_2019b}.

The parent sample from which the subsequent sub-samples are drawn was defined by applying cuts to the APOGEE-2 DR17 catalogue of data for 733,900 stars. Only stars with spectral parameters determined with confidence were considered, i.e., those with parameter ASPCAPFLAG\footnote{The ASPCAPFLAG bitmask indicates issues associated with the ASPCAP fits, which could possibly raise the uncertainties in the stellar parameters and/or elemental abundance derivations.  For additional information, the reader is referred to the APOGEE DR17 {\tt allStar} data model at {\tt https://data.sdss.org/datamodel/}} = 0. Next, we selected stars with 3000~$\leq~T_\textrm{eff}~\leq~6000$~K and $\log g < 3.6$, and whose combined DR17 spectra have S/N$>$70, in order to limit the sample to stars with reliable elemental abundances while maximising the number of Terzan~5 candidate members (see Section~\ref{sec:ter5}). At high $T_{\rm eff}$ absorption lines tend to become too weak and at low $T_{\rm eff}$ model atmospheres too uncertain for accurate abundance determination. The high $\log g$ cut is imposed to eliminate contamination by foreground dwarfs.

\cite{Massari_2014} presented an analysis of a large sample of Terzan~5 and bulge stars, based on ESO/VLT UVES and FLAMES data.  There are four stars in common between that work and the APOGEE DR17 catalogue, only three of which are considered in this work to be Terzan~5 candidate members (see Section~\ref{sec:ter5}).  We find that iron abundances from ASPCAP are lower than those by \cite{Massari_2014} by 0.4-0.6~dex.  This discrepancy could be possibly addressed by consideration of a third data set, such as the abundances derived by \cite{Origlia2011} on the basis of Keck/NIRSPEC data for 33 giants, but unfortunately there are no stars in common between our sample and that study.  We speculate that this sizeable discrepancy may result from systematic effects impacting different analysis methods at $T_{\rm eff}\simless4000$~K.  Indeed, \cite{Massari_2014} did not consider stars in that $T_{\rm eff}$ range when constructing their Terzan~5 metallicity distribution function, in order to minimise the impact of TiO bands on their metallicity determinations. However, while keeping the above caveat in mind, we point out that consideration of this zero-point difference at face value would place our sample stars within the peak of the \cite{Massari_2014} Terzan~5 MDF, which makes our sample representative of the bulk of the stellar populations in that system.  

Most importantly, such systematic departures from results from other studies should not affect our conclusions in a substantial way.  It is well understood that the abundance analysis of cool giants are notoriously uncertain.   Most of the uncertainties have a systematic nature, stemming from limitations in model atmospheres and the modelling of molecular lines (due to, e.g., line list incompleteness and $\log gf$ errors).  Our strictly differential approach makes the results in this work less prone to be significantly affected by such systematic effects. 

The evolution of APOGEE abundances along various data releases has been documented in previous publications \citep{Holtzman_2015,Holtzman2018,Jonsson2020}, so we refer the reader to those papers for details.  Nevertheless, we contrasted the stellar parameters and elemental abundances from DR17 with those from \cite{Schiavon_2017b}, which were based on DR12.  The differences are negligible, typically of the order of a few 10~K, $\sim$0.2 and $\sim$0.1 dex in $T_{\rm eff}$, $\log g$, and elemental abundances.

\begin{figure}
\begin{center}
\graphicspath{ {images/}}
\includegraphics[width=0.9\columnwidth]{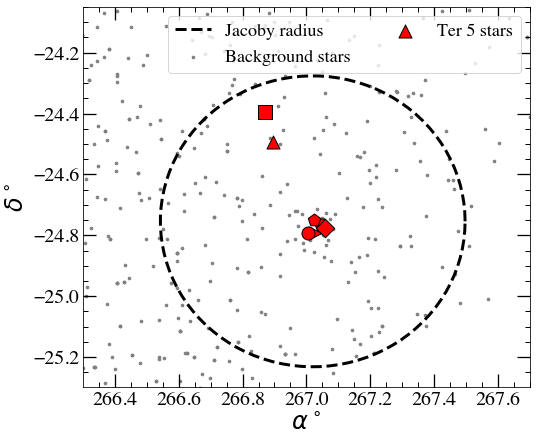}
\caption{Right ascension $\alpha$ and declination $\delta$ (in degrees) of candidate Terzan~5 stars (red shapes) plotted over the background stars (grey dots), with the cluster Jacoby radius $r_J$ (black dashed line) displayed as a reference. Five candidate members are clustered around the centre.}\label{Fig1}
\end{center}
\end{figure}

% \afterpage

%Stars with too hot effective temperatures provide spectral lines that are too weak to determine abundances and for those that are too cold, model atmospheres used in the analysis of abundances are too uncertain. We thus additionally apply the criterion: $3000 < T_\textrm{eff} < 6000$ K. Subsequently, we limit the signal-to-noise of stars to improve the data precision of the sample, and surface gravity to remove dwarf stars with characteristically large values, applying the criteria: SNR $> 70$ and $\log {g} < 3.6$.

\begin{table*}
    \centering
    \caption{Sample of candidate stars measured for Terzan~5, in order of increasing right ascension.}
    \begin{tabular}{@{}cccccccccc@{}}
    \hline\hline
    APOGEE ID & $T_\textrm{eff}$ (K) & $\log {g}$ & SNR & RV (\textrm{km/s}) & r$^\prime$ & [Fe/H] & [C/Fe] & [N/Fe] & [O/Fe]\\
    \hline%
2M17472880-2423378 & 3964 & 0.94 & 141 & $-79.4$ & 23.1 & $-0.75 \pm 0.01$ & $-0.05 \pm 0.02$ & $+0.16 \pm 0.02$ & $+0.25 \pm 0.01$ \\
2M17473477-2429395 & 4085 & 1.81 & 158 & $-80.6$ & 17.0 & $+0.26 \pm 0.01$ & $+0.05 \pm 0.01$ & $+0.29 \pm 0.01$ & $+0.06 \pm 0.01$ \\
2M17480088-2447295 & 3992 & 1.13 & 268 & $-99.2$ & 2.41 & $-0.60 \pm 0.01$ & $-0.35 \pm 0.01$ & $+1.26 \pm 0.01$ & $+0.26 \pm 0.01$ \\
2M17480576-2445000 & 4026 & 1.24 & 95  & $-76.9$ & 0.33 & $-0.63 \pm 0.02$ & $+0.04 \pm 0.02$ & $+0.78 \pm 0.02$ & $+0.30 \pm 0.02$ \\
2M17480668-2447374 & 3973 & 1.13 & 189 & $-89.9$ & 2.41 & $-0.61 \pm 0.01$ & $-0.26 \pm 0.01$ & $+1.05 \pm 0.01$ & $+0.27 \pm 0.01$ \\
2M17480857-2446033 & 3755 & 0.76 & 173 & $-64.2$ & 1.18 & $-0.73 \pm 0.01$ & $+0.17 \pm 0.01$ & $+0.46 \pm 0.02$ & $+0.32 \pm 0.01$ \\
2M17481414-2446299 & 3632 & 0.91 & 109 & $-76.0$ & 2.46 & $+0.07 \pm 0.01$ & $+0.02 \pm 0.01$ & $+0.07 \pm 0.01$ & $+0.07 \pm 0.01$ \\
    \hline\hline
    \end{tabular}
\label{tab2}
\end{table*}

\begin{table*}
    \centering
    \begin{tabular}{@{}ccccccc@{}}
    \hline\hline
    [Mg/Fe] & [Al/Fe] & [Si/Fe] & [S/Fe] & [Ca/Fe] & [Mn/Fe] & [Co/Fe]\\
    \hline%
$+0.30 \pm 0.02$ & $-0.05 \pm 0.03$ & $+0.17 \pm 0.02$ & $+0.50 \pm 0.08$ & $+0.11 \pm 0.02$ & $-0.07 \pm 0.03$ & $+0.09 \pm 0.05$ \\
$+0.02 \pm 0.01$ & $-0.03 \pm 0.02$ & $-0.03 \pm 0.01$ & $+0.01 \pm 0.03$ & $-0.05 \pm 0.01$ & $+0.31 \pm 0.01$ & $+0.12 \pm 0.03$ \\
$+0.25 \pm 0.01$ & $+0.53 \pm 0.02$ & $+0.17 \pm 0.01$ & $+0.21 \pm 0.05$ & $+0.32 \pm 0.02$ & $-0.07 \pm 0.02$ & $+0.39 \pm 0.03$ \\
$+0.30 \pm 0.02$ & $+0.16 \pm 0.04$ & $+0.16 \pm 0.02$ & $+0.41 \pm 0.09$ & $+0.16 \pm 0.03$ & $-0.04 \pm 0.03$ & $+0.04 \pm 0.07$ \\
$+0.25 \pm 0.01$ & $+0.18 \pm 0.02$ & $+0.16 \pm 0.01$ & $+0.23 \pm 0.05$ & $+0.11 \pm 0.02$ & $-0.05 \pm 0.02$ & $+0.35 \pm 0.03$ \\
$+0.28 \pm 0.02$ & -                & $+0.19 \pm 0.02$ & -                & $+0.08 \pm 0.02$ & -                & $+0.26 \pm 0.03$ \\
$+0.01 \pm 0.01$ & -                & $-0.04 \pm 0.02$ & -                & $-0.11 \pm 0.02$ & -                & -                \\
    \hline\hline
    \end{tabular}
\end{table*}

\subsection{Bulge Sample} \label{sec:bulge}
To create the sub-sample defining stars from the Galactic bulge, the Galactocentric distance $R_\textrm{GC}$ for each star was determined using Galactic longitude \emph{l}, latitude \emph{b}, \texttt{astroNN} distance $d$, and distance error $d_{\textrm{err}}$, and assuming a distance of 8 kpc between the Sun and the Galactic centre \citep{Utkin_2020}. An additional cut was applied to remove stars with large distance errors, thus a mask was applied to the parent sample for stars matching the criteria:

\begin{enumerate}
    \item Galactocentric distance $R_\textrm{GC} < 4$ kpc,
    \item Fractional distance uncertainty $\sigma_d/d < 0.2$.
\end{enumerate}

Furthermore, stars for which compositions in Mg, C, N, O, and Si could not be determined by ASPCAP were removed, along with 99 members of bulge clusters from the APOGEE GC value added catalog from Schiavon et al. (2021, in prep.), resulting in a final sample of 21,052 bulge stars.

\subsection{Terzan~5 Candidates} \label{sec:ter5}
Terzan~5 candidates were selected on the basis of angular distance from the cluster centre, radial velocity, and proper motion.  Central coordinates adopted for Terzan~5 were $\alpha = 17^\textrm{h}48^\textrm{m}4.80^\textrm{s}$ and $\delta = -24^\circ46^{\prime}45^{\prime\prime}$, taken from \citet{harris2010new}. Relevant values describing Terzan~5 are summarised in Table \ref{tab1}. Stars were considered to be candidate members of Terzan~5 if they are located within the cluster Jacoby radius and if their radial velocities differed from the mean cluster radial velocity by no more than two times the cluster's velocity dispersion.  In addition, we adopted their cluster mean radial velocity RV$_\textrm{Ter 5} = -82.57$ km/s and dispersion $\sigma_\textrm{RV} = 15.5$ km/s. Additional criteria are based on {\it Gaia} EDR3 \citep{Gaia_2021_sum, Gaia_2021_phot} proper motions (PM). Terzan~5 candidate members are considered to be those whose proper motions do not differ from the mean value of the cluster \citep{Baumgardt_2021} by more than five times the cluster mean PM dispersion, $\sigma_\textrm{PM}$ (Schiavon et al. 2022, in prep.). Hence, a mask was applied to the parent sample for stars matching the following criteria:

\begin{enumerate}
    \item Distance from cluster centre $r < 25.37^\prime$,
    \item Radial velocity in the range $-113.57 <$ RV $< -51.57$ km/s,
    \item $\delta$pm $< 2.7$ mas/yr.
\end{enumerate}

\noindent where $r_J=25.37^\prime$ is the Jacoby radius of Terzan~5, taken from \citet{Baumgardt_2021}\footnote{\url{https://people.smp.uq.edu.au/HolgerBaumgardt/globular/} (May 2021 - 3$^\textrm{rd}$ version)}, $\delta$pm is the proper motion residual relative to that of Terzan~5.  Due to the large spread in [Fe/H] for Terzan~5, no metallicity criterion was adopted. 

A note of caution is required in regards to the adoption of the Jacoby radius of Terzan~5.  Measurements of tidal radii of globular clusters are notoriously uncertain, a problem that is further exacerbated in crowded regions such as the Galactic bulge.  To illustrate this uncertainty we refer to the work by \cite{Lanzoni_2010}, who obtained a much smaller tidal radius of $4.^\prime6$ from fitting a \cite{King_1966} profile to a combination of ground-based and {\it Hubble Space Telescope} data.  The much larger Jacoby radius from \citet{Baumgardt_2021} results from a dynamical calculation matching the cluster's stellar density and velocity dispersion profiles (derived from a combination of literature sources and Gaia eDR3).  The Jacoby radius depends on the cluster's mass and orbit.  By definition, Jacoby radius and King tidal radius do not necessarily agree, as clusters do not follow a King profile at distances of the order of the Jacoby radius.  Nevertheless, the uncertainty in the Jacoby radius of Terzan~5 is non-negligible, because the cluster's orbit is not well known (H. Baumgardt, 2022, priv. comm.).

Application of the above selection criteria initially resulted in the consideration of 9 candidate stars, 5 of which are located within the \cite{Lanzoni_2010} tidal radius.  Out of the remaining four stars, two differ in RV from the mean cluster value by more than $1\sigma_{\rm RV}$ while being located at angular distances comparable to $r_J$.  Since the cluster velocity dispersion is known to drop substantially at such large distances\footnote{\url{ https://people.smp.uq.edu.au/HolgerBaumgardt/globular/veldis.html}}, we decided to not consider these two stars as candidate members.  This resulted in a sample of 7 candidate stars which are adopted in the subsequent analysis. Their properties are summarised in Table~\ref{tab2}. Candidate stars can also be seen plotted on a graph of celestial coordinates in Figure~\ref{Fig1}, with all other stars in APOGEE-2 DR17 shown in the background.

In addition to that fiducial sample of seven Terzan~5 candidate stars, we assess the impact of our Jacoby radius selection by re-running our analysis on the more stringent candidate member sample of five stars located within the \citet{Lanzoni_2010} tidal radius of the cluster centre.  Our results are essentially unchanged, as discussed in Section~\ref{sec:compare}, where we provide a summary of the  results for both candidate member samples.  Finally, We note that star 2M17475169-2443153 was considered a possible Terzan~5 member by \cite{Schiavon_2017b}, but due to its discrepant Gaia-based proper motion it is not included in our sample.

% \FloatBarrier
\section{Analysis and Results}
\label{sec:analysisresults} % used for referring to this section from elsewhere

In possession of a vetted sample of Terzan~5 members, we proceed to compare the detailed abundance pattern of that cluster with that of the Galactic bulge field.  In this Section we quantify the similarity of these abundance patterns, in order to test the hypothesis that the progenitor of Terzan~5 is a major contributor to the mass of the bulge stellar populations.

\subsection{Terzan~5 versus bulge chemistry} \label{sec:compare}

\begin{figure}
\centering
\includegraphics[width=0.85\linewidth]{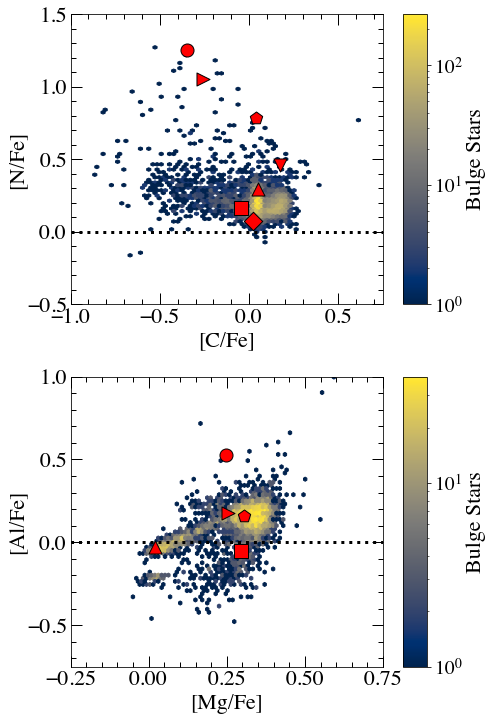}
\caption{[N/Fe]-[C/Fe] (top panel) and [Al/Fe]-[Mg/Fe] (bottom panel) distributions of bulge field stars (hex-bin data points described by the colour bar) and Terzan~5 stars (red shapes), where both planes include the three N-rich Terzan~5 stars subsequently withdrawn from the sample for only abundances affected by the MP phenomenon. The number of Terzan~5 stars displayed in each panel differs due to the absence of two [Al/Fe] ASPCAP abundances. Each panel includes the solar abundance as a black dotted line.}\label{Fig2}
\end{figure}

Abundance ratios adopted in our analysis include the following elements, chosen as they are able to be reliably determined by ASPCAP: C, N, O, Mg, Al, Si, S, Ca, Mn, and Co. Prior to carrying out comparisons of the detailed chemical compositions of Terzan~5 with those of their field bulge counterparts, we need to refine the sample used for comparisons in chemical spaces using abundances for elements that are affected by the MP phenomenon in GCs \citep[e.g.,][]{Bastian_2020, Renzini_2008}. In particular, light elements 
%involved in proton-capture processes 
such as C, N, O, Mg, and Al exhibit important star-to-star abundance variations that would severely bias the comparison with the field population.  In particular, \cite{Schiavon_2017b} showed that this phenomenon is present amongst Terzan~5 stars.  To account for this effect, we remove from the comparisons with the field sample any Terzan~5 stars with abundances typical of the so-called ``second-generation'' stars. They can be easily identified in Figure \ref{Fig2}, where the distribution of the two samples in both the [N/Fe]-[C/Fe] and [Al/Fe]-[Mg/Fe] planes are shown. We adopt a threshold of [N/Fe] $= +0.5$, above which stars are considered to have abnormal abundance patterns. As a result, the Terzan~5 sample is reduced to 4 stars for the affected abundances (C, N, O, Mg, and Al), whereas all 7 stars are adopted in the comparisons involving all other elements.

\begin{figure*}
\centering
\includegraphics[width=0.77\linewidth]{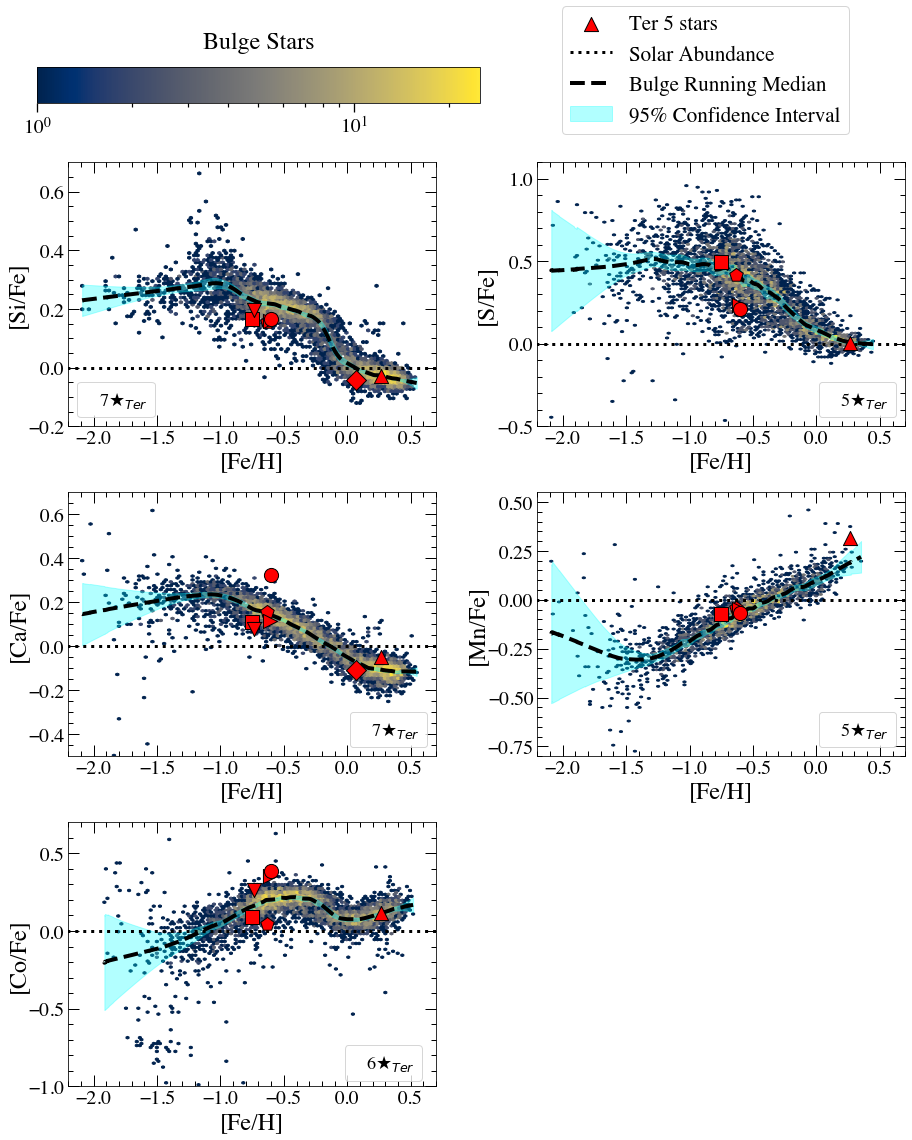}
\caption{[X/Fe]-[Fe/H] distributions of abundances not affected by the MP phenomenon, with hex-bin data points described by the colour bar to indicate bulge field stars (in bins of 100 and 80 for [Fe/H] and [X/Fe] abundance, respectively) and Terzan~5 stars (red triangles) with applicable abundances. Each panel includes the solar metallicity (black dotted line), its respective bulge sample running median (black dashed line), and the number of Terzan~5 stars ($\bigstar_\textrm{Ter}$) with acceptable abundances shown.}\label{Fig3}
\end{figure*}

\begin{figure*}
\centering
\includegraphics[width=0.77\linewidth]{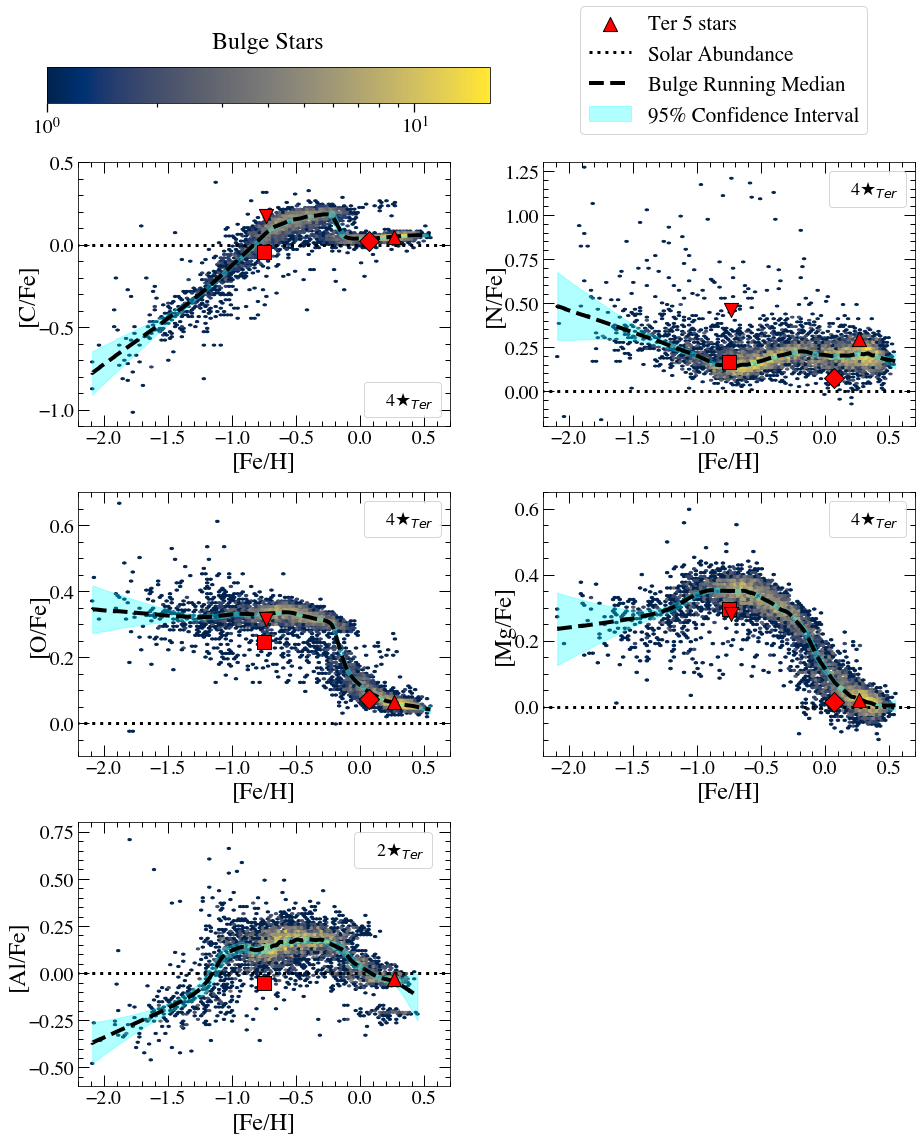}
\caption{[X/Fe]-[Fe/H] distributions of abundances affected by the MP phenomenon, with hex-bin data points described by the colour bar to indicate bulge field stars (in bins of 100 and 80 for [Fe/H] and [X/Fe] abundance, respectively) and the sample of Terzan~5 stars (red triangles) with applicable abundances, reduced by removing the three most N-rich stars. Each panel includes the solar metallicity (black dotted line), its respective bulge sample running median (black dashed line), and the number of Terzan~5 stars ($\bigstar_\textrm{Ter}$) with acceptable abundances shown.}\label{Fig4}
\end{figure*}

The resulting Terzan~5 stars are contrasted with the bulge sample in various chemical planes in Figures~\ref{Fig3} and \ref{Fig4}, where the former/latter include elements that are/are not affected by the MP phenomenon. In both sets of plots, the 2D histogram indicates the bulge sample within a narrow range in $\log g$ ($\pm$ 0.25~dex) around the mean of Terzan~5 for the reasons discussed below.  The sample of Terzan~5 member candidates is shown as red symbols, which are assigned consistently to each star for easy identification across multiple plots.  To guide the eye, the running median of the bulge sample is indicated by the dashed lines -- determined using the {\it statsmodels} locally
weighted scatter-plot smoothing (LOWESS\footnote{\url{https://www.statsmodels.org/devel/generated/statsmodels.nonparametric.smoothers_lowess.lowess.html}}) algorithm \citep{Cleveland_1979}, weighted to a fraction of 0.07 of the data surrounding each data point and iterated 3 times. We estimated the 95\% confidence interval (cyan shading) via bootstrapping 25\% of the data 100 times and estimating the resulting spread. Visual inspection suggests that there are important differences between Terzan~5 and the bulge field for $\alpha$-elements such as Si, Ca, O, and Mg, as well as Fe-peak element Mn. For other elements, differences are likely absent, or present but more subtle.  

However suggestive the comparisons displayed in Figures~\ref{Fig3} and \ref{Fig4} might seem, we need a quantitative estimate of the chemical differences between Terzan~5 and the bulge. It is crucial that such differences be quantified in a statistically robust fashion. To achieve this goal, we calculate the offset of the Terzan~5 stars from the bulge sample in various abundance planes. For any given element X, we define the quantity $\rho_X$ as follows:
\begin{equation}
{\bf \rho_X = {\rm median} \left(\frac{ {\rm [X/Fe]}^\textrm{Ter~5}_i - {\rm [X/Fe]}^\textrm{Bulge}_i}{\sqrt{\sigma^2{\rm [X/Fe]}^{\rm Ter~5}_i + \sigma^2{\rm [X/Fe]}^{\rm Bulge}_i} }\right)}
\end{equation}

\noindent where [X/Fe]$^{\rm Ter~5}_i$ and $\sigma$[X/Fe]$^{\rm Ter~5}_i$ are the abundance ratio and error of element X in Terzan~5 star $i$, [X/Fe]$^{\rm Bulge}_i$ is the median of [X/Fe] calculated for a sub-sample of the bulge field selected to narrowly match the [Fe/H] and $\log g$ values of Terzan~5 star $i$, and $\sigma$[X/Fe]$^{\rm Bulge}_i$ the error in the median.  
%{\red The $\rho$ statistic is akin to  the likelihood of Terzan~5 being similar to the bulge in a given abundance plane.}

For each Terzan~5 star a sub-sample of the bulge stars of same [Fe/H] must be selected for the calculation of $\rho_X$.  In addition to selecting field stars with similar [Fe/H] as that of Terzan~5, we need to control for $\log g$ so as to minimise the impact of systematics in the ASPCAP abundance determinations.  \cite{Weinberg_2021} showed that such systematics are responsible for important artificial variations in elemental abundance as a function of position along the giant branch \citep[for a detailed discussion see also][Horta et al. 2022, in prep., and Kisku et al. 2022, in prep.]{Eilers2021}
. Thus, the bulge field stars selected for the comparison differed by no more than 0.1~dex in [Fe/H] and 0.25~dex in $\log g$ from each Terzan~5 candidate members.

The number of Terzan~5 stars considered in the calculation, $n$, was equal to 4 for elements affected by the MP phenomenon and 7 for other elements, though this changed depending on whether or not ASPCAP could provide an acceptable elemental abundance for each star.  In the case of Al, for instance, abundances are available for only 2 Terzan~5 stars not affected by the multiple populations phenomenon.

In order to gain a better grasp of the significance of $\rho_X$ obtained for Terzan~5, we estimate the values that would be expected in the case that Terzan~5's chemistry is identical to that of the bulge.  That was achieved through a bootstrapping technique, where $\rho_X$ was calculated for each element X for 1,000  Terzan~5-sized random samples drawn from the bulge population by picking stars with [Fe/H] and $\log g$ similar to those of our Terzan~5 sample, with replacement.  Thus, for each abundance ratio we obtain 1,000 random samples of a maximum of 7 stars from the bulge population by selecting stars within a narrow range of [Fe/H] and $\log g$ around the candidate Terzan~5 stars.  Mean sizes of the comparison bulge sample selected around candidate cluster members range from 700 to 911, depending on the element and the star, with the minimum size of 340 in Mn, for the most metal-poor Terzan~5 star.  In order to preempt artificial differences being induced by systematic effects in the ASPCAP abundances, we proceeded in precisely the same way for each random sample. Therefore, for each star $i$ of the random sample, [X/Fe]$^{\rm Bulge}_i$ was the median value from a bulge field sub-sample selected to differ in [Fe/H] and $\log g$ from the $i$ star by no more than 0.1 and 0.25~dex, respectively.

The above procedure generates $\rho_X$ distributions based on those random samples for each elemental abundance ratio [X/Fe], which can then be compared with the $\rho_X$ obtained from comparison of the Terzan~5 sample itself with the bulge field samples. If the  abundances of element X in the Terzan~5 stars differ significantly from that of their bulge field counterparts, $\rho_X$ obtained from the Terzan~5 sample should differ from that of the median of the random bulge samples in a statistically significant way.

The $\rho_X$ distributions of abundances not affected by the MP phenomenon, and those that are, are shown in Figures~\ref{Fig5} and \ref{Fig6}, respectively. The median $\rho_X$ of the random samples is indicated by the cyan dashed lines and that for the Terzan~5 sample as a red dashed line. The light and grey shades indicate the regions within 1 and 2$\sigma$ away from the median of the $\rho_X$ distributions, and included in each panel is the factor of sigma that the $\rho_X$ medians differ by.  We find that Terzan~5 differs from the bulge at least at the 1$\sigma$ level in all abundances except for S, Co, and N, and by 2$\sigma$ or more in Si and Mg. For Ca, Mn, C, O, and Al, the two systems differ at a level between 1$\sigma$ and 2$\sigma$. The element for which the difference is the most significant is Si, at $\sim 4\sigma$.  Perhaps most significantly, $\alpha$-elements Mg, Ca, O, and Si are all consistently depressed in Terzan~5 relative to the bulge, whereas the most reliable Fe-peak element in our sample, Mn, is significantly enhanced in Terzan~5.

% The sample of Terzan~5 stars used by \citet{Nataf_2019} consists of 9 stars, 7 of which are also included in our fiducial sample. In order to demonstrate that our results are not strongly sensitive to the details of our Terzan~5 candidate sample, we reran the above exercise limiting our Terzan~5 sample to the list of candidates considered by \citet{Nataf_2019}. Results from this run show a difference between Terzan~5 and the bulge of 2$\sigma$ or more across a much wider range of elements: Si, Ca, Co, N, O, Mg, and Al. These are shown in Figures~\ref{FigA1} and  \ref{FigA2}.

\begin{figure}
\centering
\includegraphics[width=0.95\linewidth]{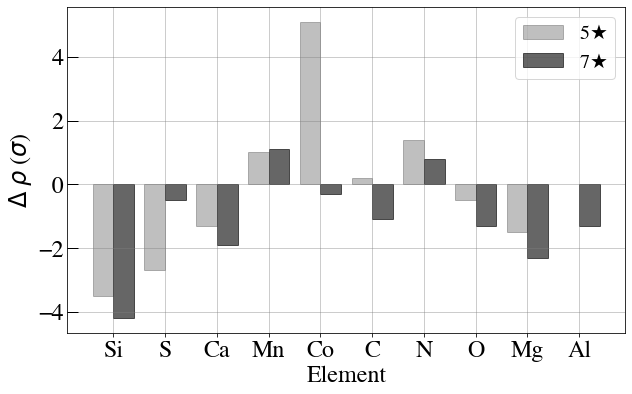}
\caption{Results of the random sampling technique for both the most stringent (light shading) and our adopted sample (dark shading) of Terzan~5. Shown is the separation $\Delta \rho$, in units of standard deviations of the bulge $\rho$-distribution, between the median of the randomly sampled bulge $\rho$-distribution and Terzan~5 $\rho$ for each element analysed.  The result for Al is omitted for the 5-star sample due to the absence of available Terzan~5 abundances. }\label{Figbar}
\end{figure}

\begin{figure*}
\centering
\includegraphics[width=0.71\linewidth]{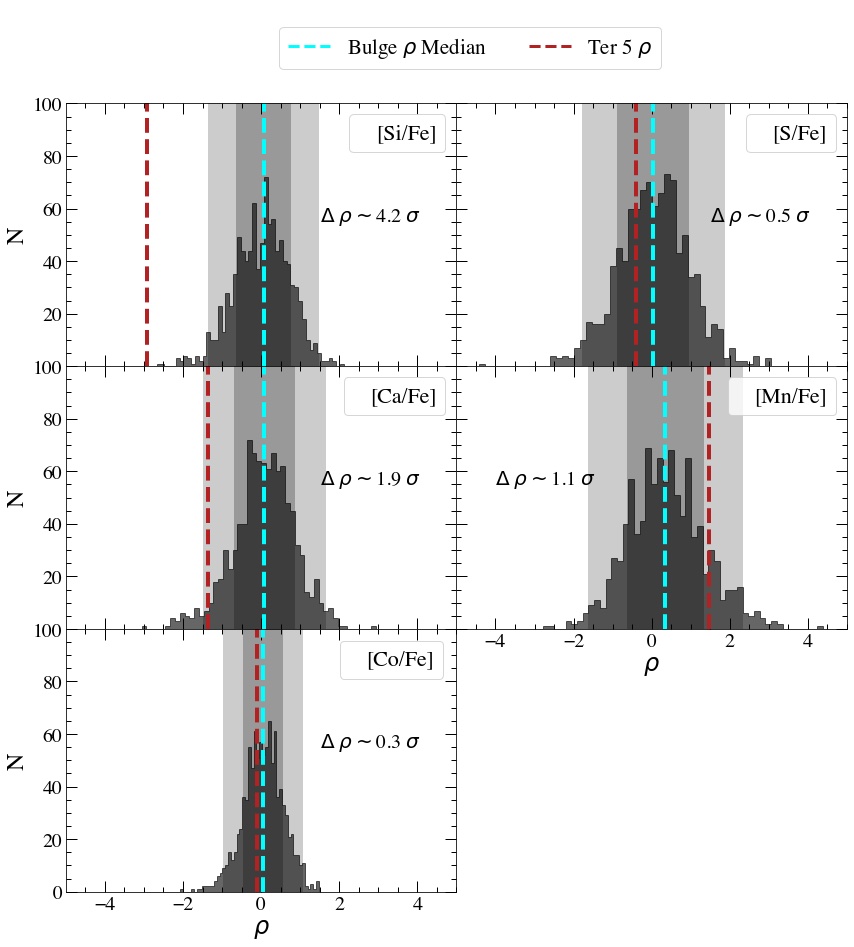}
\caption{Histograms of the $\rho$-distributions of the selected random bulge samples for abundances {\it not} affected by the MP phenomenon. Each panel includes the corresponding median of the randomly sampled bulge $\rho$-distribution (cyan dashed line) -- with light and dark shading indicating its 1$\sigma$ and 2$\sigma$ error, respectively -- and Terzan~5 $\rho$ (red dashed line), along with their separation, $\Delta \rho$.}\label{Fig5}
\end{figure*}

\begin{figure*}
\centering
\includegraphics[width=0.71\linewidth]{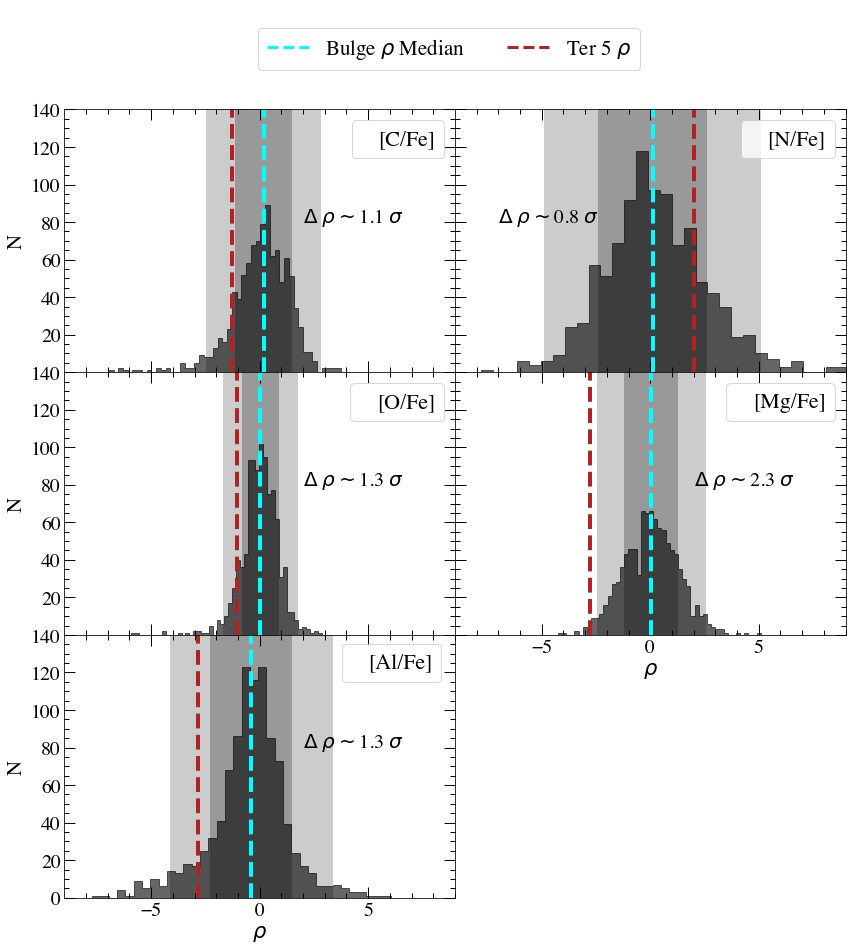}
\caption{Histograms of the $\rho$-distributions of the selected random bulge samples for abundances affected by the MP phenomenon. Each panel includes the corresponding median of the randomly sampled bulge $\rho$-distribution (cyan dashed line) -- with light and dark shading indicating its 1$\sigma$ and 2$\sigma$ error, respectively -- and Terzan~5 $\rho$ (red dashed line), along with their separation, $\Delta \rho$.}\label{Fig6}
\end{figure*}

We further checked the sensitivity of our results to the Terzan~5 sample selection by running our analysis on the alternative, more stringently selected Terzan~5 candidate member sample of 5 stars, by removing stars {\tt 2M17472880-2423378} and {\tt 2M17473477-2429395} from the sample (represented respectively by the square and right-side-up triangle shapes in all Figures). The summary of the results from the random sampling technique are provided in Table \ref{tab3} and Figure \ref{Figbar}. The outcome of this exercise is a broad confirmation of the results obtained based on our fiducial Terzan~5 sample.
% , although the number of elements according to which Terzan~5 is a mismatch to the bulge distribution by 2$\sigma$ or more reduces to two when the most stringently selected sample is adopted.
Comparisons between $\rho$-distributions and mean candidate member abundances for the stricter sample are displayed in  Figures~\ref{FigA1} and \ref{FigA2} of the Appendix.  All the results obtained on the basis of the fiducial sample are confirmed, with a slightly decreased statistical significance.  The exceptions are cobalt, sulfur, and, to a lesser extent, nitrogen.  Figures~\ref{FigA1} and \ref{FigA2} show that the differences in $\rho_X$ between Terzan~5 and the median bulge jumps to statistically significant values for these elements when shifting to the stricter sample.  Sulfur and cobalt abundances in APOGEE have only moderate precision due to being based on few lines \citep[two for S, one for Co, see][]{Jonsson2020}.  It is interesting that, when based on a stricter sample, abundances for both elements show the same behaviour as those of elements from the same nucleosynthetic family ($\alpha$ in the case of sulfur, Fe-peak in the case of cobalt).  Given the large variance of the [S/Fe] and [Co/Fe] values for Terzan~5 stars and the small sample, we do not place great confidence in this result.

%Most of the cluster's abundances driving the offset from the bulge in this sample are Co-richer than the median of the bulge, therefore it is definitely reasonable for the offset to increase in magnitude by such an amount when this abundance is removed, reassuring us that the method proves acceptable.}

\begin{table}
    \centering
    \caption{Results of the random sampling technique for both the most stringent and our adopted sample of Terzan~5. Column definitions are as follows: (1) Number of considered cluster candidates; (2) elements exhibiting a $\rho$ separation of $1\sigma \leq \Delta\rho < 2\sigma$; (3) elements exhibiting a $\rho$ separation of $\Delta\rho \geq 2\sigma$. Results of our adopted sample are given in the shaded row.}
    \begin{tabular}{@{}ccc@{}}
    \hline\hline
    (1) & (2) & (3) \\
    \hline
    N$_\textrm{Ter 5}$ & X$_{1\sigma \leq \Delta\rho < 2\sigma}$ & X$_{\Delta\rho \geq 2\sigma}$\\
    \hline%
5 & Ca,Mn,N,Mg & Si,S,Co \\
\rowcolor{Gainsboro!60}
7 & Ca,Mn,C,O,Al & Si,Mg \\
    \hline\hline
    \end{tabular}
\label{tab3}
\end{table}

%{\red Interestingly, the number of elements for which Terzan~5 differs from the bulge by 2$\sigma$ or more is maximal for our fiducial sample of 7 stars, suggesting that our preferred selection criteria strike an optimal compromise between maximising statistical significance and minimising field star contamination. Nonetheless, a larger sample of {\it bona fide} cluster members is required to check the reality of our results.}
% , especially as results obtained using the same number of stars, but for the sample considered in \citet{Nataf_2019}, display a much larger chemical difference between Terzan~5 and bulge field than when employing our fiducial sample.

In summary, we conclude that our data are consistent with Terzan~5 and the bulge field being chemically distinct, with $\alpha$ elements Si, Ca, Mg, and O being depressed, and Fe-peak element Mn enhanced, in Terzan~5 in a statistically significant way.  In the next sections we discuss how our results constrain existing models for the nature of Terzan~5.

%This is indicative of the need for obtaining a larger sample of Terzan~5 stars in future work, to add statistical robustness to our result. 

%While on the other hand, this supports the use of the adopted sample which is neither too strict, which would devalue the efficiency of the sampling technique, or too lenient, which would obtain a larger fraction of bulge contaminants.

\section{The Nature of Terzan~5}
\label{sec:nature}

In this Section we examine the implications of our results to scenarios proposed in the literature to explain the properties of Terzan~5.  We focus on three different hypotheses: (1) Terzan~5 is the leftover of a dark-matter dominated accreted satellite; (2) Terzan~5 results from the evolution of a massive clump resulting from disc instabilities at high redshift; and (3) Terzan~5 is an old globular cluster rejuvenated by star formation based on gas resulting from accretion due to encounters with giant molecular clouds.

\subsection{Terzan~5 as the remnant of a dark matter dominated system}

We concluded in the previous Section that the abundance patterns of Terzan~5 and the bulge field differ in a statistically significant way.  At face value, this result implies that the progenitor of Terzan~5 cannot have made an important contribution to the stellar mass budget of the bulge, otherwise their abundance patterns would be similar.  This result is apparently at odds with the qualitative agreement found in previous work \citep[e.g.,][]{Ferraro_2016} between the distribution of their stars in the $\alpha$-Fe plane, particularly in regards to the position of the knee, which is an indicator of the stellar mass of the system \citep[][Mason et al.\ 2022, in prep.]{Tolstoy_2009}.  

It is generally believed that it takes a massive system, dark-matter dominated, to foster the type of chemical evolution responsible for a stellar distribution on the $\alpha$-Fe plane such as seen in Terzan~5 and the Galactic bulge. Assuming that [Fe/H]$_\textrm{knee}$ is an unequivocal estimator of stellar mass, we are thus left with a conundrum: the progenitor of Terzan~5 may have had a stellar mass that is comparable to that of the bulge, which implies that it obviously lost most of its stars to the bulge.  Yet their abundance patterns differ to a reasonable degree of statistical significance.

%[[not clear that we should state this here]] As we will see, Terzan~5 probably formed the bulge with the aid of many other progenitors.

However, empirical evidence shows that the scaling relation between [Fe/H]$_\textrm{knee}$ and M$_\star$ exhibits notable scatter. For example, despite the Fornax dSph being $\sim$10x brighter than the Sculptor dSph, they have similar [Fe/H]$_\textrm{knee}$ \citep{Hendricks_2014}.  Moreover, the Large and Small Magellanic Clouds exhibit low  [Fe/H]$_\textrm{knee}$ values for their large masses \citep{Nidever_2020}. \cite{deBoer_2014} proposed that the dependence of  [Fe/H]$_\textrm{knee}$ on stellar mass is affected by details of its star-formation history, which in turn is dependent on its total mass.  Given these considerations, a reassessment of the mass of the putative progenitor of Terzan~5 on the basis of state-of-the-art theoretical predictions is in order.

\subsubsection{Progenitor mass estimate using the EAGLE simulations}

In a new theoretical study, Mason et al.\ (2022, in prep.) show that [Fe/H]$_\textrm{knee}$ is not solely determined by stellar mass, M$_\star$, but is also affected by details of its star-formation history \citep[see also][]{Andrews2017}.  In this Section we use these theoretical predictions for the relation between [Fe/H]$_\textrm{knee}$ and M$_\star$ in order to estimate the possible range of masses of the progenitor of Terzan~5.

\begin{figure}
\centering
\includegraphics[width=0.90\linewidth]{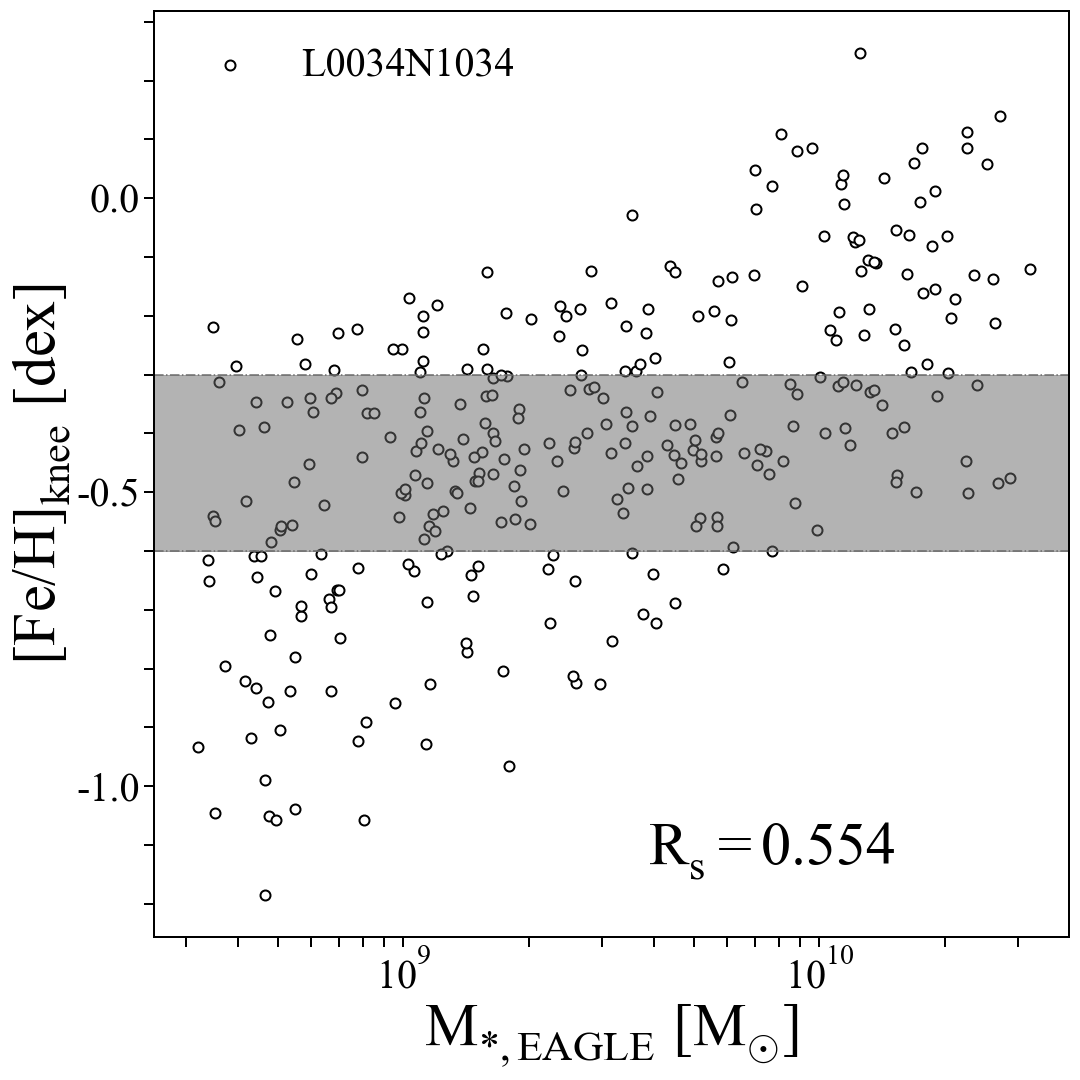}
\caption{The distribution of [Fe/H]$_{\rm{knee}}$, as a function of $\rm{M}_{\star}$, for simulated galaxies with knees in the L034N1034-RECAL volume of the EAGLE simulations. R$_\textrm{s}$ is the scaling coefficient, and the shading indicates the Terzan~5 progenitor mass range given its estimated [Fe/H]$_{\rm{knee}}$.}\label{fig5 - knee}
\end{figure}

For this purpose we employ predictions based on the Virgo Consortium's Evolution and Assembly of GaLaxies and their Environments (EAGLE) suite of numerical hydrodynamical simulations \citep{schaye_2015, crain_2015}, which follow the formation of galaxies evolving in cosmologically representative volumes of a standard $\Lambda$-CDM model of the Universe.  Figure~\ref{fig5 - knee} shows the theoretical prediction from Mason et al.\ (2022, in prep.) for that relation, based on the analysis of data for galaxy populations from a high-resolution volume of the EAGLE simulations (L034N1034-RECAL), which evolves $1034^3$ dark matter and gas particles in a volume comprising a periodic cube with length 34 cMpc on a side. The EAGLE-based theoretical prediction broadly confirms the expectations in the literature for a  monotonically increasing relationship between [Fe/H]$_\textrm{knee}$ and M$_\star$.  In addition, they predict a  significant scatter in that relation, in qualitatively good agreement with the observations.  For more details, see Mason et al.\ (2022, in prep.). 

We estimate [Fe/H]$_\textrm{knee}$ for Terzan~5 from the distribution of mean [$\alpha$/Fe] as a function of [Fe/H], where the elements entering the mean were Si, S, Ca, and O. Considering the uncertainties, we estimate $-0.6\simless$ [Fe/H]$_\textrm{knee}\simless-0.3$ for Terzan~5. Considering that range of [Fe/H]$_\textrm{knee}$ (shaded area in Figure~\ref{fig5 - knee}), the predicted range of possible masses for the progenitor of Terzan~5 is therefore M$_\star \approx 3\times10^{8} - 3\times10^{10}$ M$_\odot$.
% \cite{Ferraro_2016} estimate a progenitor mass of a few times $10^9~{\rm M}_\odot$, which is well within the range allowed by the EAGLE simulations. FERRARO GAVE 10^9 MSOL AS THE MASS OF THE HOST GALAXY IN THE ACCRETED SCENARIO
Estimates of the mass of the Terzan~5 progenitor are naturally very uncertain.  Based on comparisons with other systems containing iron abundance spreads,  \citet{Lanzoni_2010} suggested a value of the order of $\sim 10^8$ M$_\odot$ and a possible lower limit of $\sim 10^7$ M$_\odot$, which places it at the minimum of our estimated range. Considering the system's history of star formation and chemical enrichment, \citet{Ferraro_2016} put forward an initial mass of at least a few times $\sim 10^8$ M$_\odot$. In view of our estimated range for the mass of the Terzan~5 progenitor, we next discuss the implications for its contribution to the stellar mass content of the bulge.

\subsubsection{Contribution to the bulge stellar mass}

The current mass of Terzan~5 is estimated to be of the order of $10^6~{\rm M}_\odot$ \citep{Lanzoni_2010}.  Given the above estimated range of masses of the cluster's progenitor, one concludes that the original Terzan~5 system could in principle have contributed somewhere between $10^8$ and $10^{10}~{\rm M}_\odot$ to the stellar mass budget of the Galactic bulge.  These numbers should be contrasted with the total stellar mass within the central few kpc of the Galactic centre, an assessment of  which was provided by \cite{Valenti_2016}, who estimated that there are $\sim~2\times10^{10}~{\rm M}_\odot$ of stars within $|b|<9.5^\circ$ and $|l|<10^\circ$. 

In view of the above numbers, we can make an educated inference of the contribution by the Terzan~5 system to the stellar mass budget of the Galactic bulge.  We first consider the most likely case, where the mass of the progenitor occupied the low end of the range allowed by the EAGLE simulations.  In that scenario, the mass of the Terzan~5 progenitor could be as much as two orders of magnitude lower than that of stellar populations within the bulge.  Such a relatively low-mass progenitor could easily have contributed its entire stellar mass to the Galactic bulge without significantly influencing the latter's mean chemical composition, which would then explain the differences reported in Section~\ref{sec:compare}.

Next, we consider the case in which the progenitor mass was considerably more massive.  In that situation, Terzan~5 would have been the nuclear cluster of a $10^9-10^{10}{\rm M}_\odot$ dwarf galaxy that was accreted to the Milky Way.  One such massive accreted system would have contributed substantially to the stellar content of the inner Galaxy which, at face value, is at odds with the chemical differences discussed in Section~\ref{sec:compare}.  One possible way of accommodating the fact that Terzan~5 has lower [$\alpha$/Fe] than the bulge field would be the existence of a chemical composition gradient in the progenitor, so that most of its stellar mass was $\alpha$-enhanced relative to the nuclear cluster. The likelihood of such a scenario can be assessed by consideration of the chemistry of existing nuclear clusters and their host galaxies.  Take, for instance, the case of the Sagittarius dwarf spheroidal (Sgr dSph) and its nuclear cluster M54.  Recently, \citet{Hayes_2020} used APOGEE DR16 to determine median values for [Mg/Fe] $\approx -0.03$ and [Si/Fe] $\approx -0.12$ for the Sgr dSph core, and slightly larger values for both the leading and trailing arms (by no more than 0.15/0.06 dex in the case of [Si/Fe]/[Mg/Fe]).  These numbers should be confronted with those obtained by  \citet{Trincado_2021}, who determined for M54 the mean abundances [Mg/Fe] $\approx +0.2$ and [Si/Fe] $\approx +0.2$. So, both the core and tidal streams of the Sgr dSph have lower [$\alpha$/Fe] than its nuclear cluster, thus going in the opposite direction of what is required to explain the chemical discrepancies between Terzan~5 and the bulge field.  

% {\cyan [[I think this evidence is much weaker than what we are doing and doesn't need to be cited]] Meanwhile, \citet{Massari_2014,Massari_2014b,Massari_2014c} conclude that the high-metallicity regime of Terzan~5 and its chemical link with the bulge means it's very unlikely for it to have been accreted, whereby the large central concentration of the metal-rich population favours a self-enrichment scenario and}

Evidence against Terzan~5 being the remnant of an accreted dwarf galaxy has been presented in the recent literature.  For instance, \citet{Prager_2017} explored the large pulsar population of Terzan~5 to find a current stellar mass-to-light ratio $M/L_{\rm V} \sim 2-3$.  They compare that number with those of ultra-faint dwarfs, which are orders of magnitude higher \cite[e.g.,][]{mcconnachie2012}, concluding that such a drastic change in stellar mass-to-light ratio is unlikely, even after accounting for tidal stripping.  In contrast, we argue that, given our range of estimates for the progenitor mass of Terzan~5, its mass-to-light ratio should be instead compared to those of more massive dwarfs. Take, for instance, the case of Fornax, whose mass falls just below the low-mass end of our progenitor mass estimate, and has a stellar mass-to-light ratio in fact comparable to that of Terzan~5 \citep[e.g.,][]{mcconnachie2012}. So, perhaps the mass-to-light ratio of Terzan~5 may not in fact be inconsistent with a dwarf satellite origin.

More recently, \citet{Pfeffer_2020} argued that Terzan~5 is unlikely to have been part of an accreted galaxy due to its hosting a 4.5~Gyr old stellar population.  That would require the occurrence of a relatively major accretion event having taken place less than 4.5~Gyr ago, for which there is no evidence, though there is evidence for a possible population of bulge stars with similar ages \citep[e.g.,][]{Bensby_2020}. 

%In addition, according to \cite{Pfeffer_2020}, such a massive dwarf galaxy would have likely hosted at least a dozen GCs and there is no report of such a significant family in the Galactic bulge with similar orbital properties as Terzan~5.

In view of our results, we cannot rule out the hypothesis that Terzan~5 is the remnant of an accreted dark matter-dominated system.  However, the chemical differences between Terzan~5 and its bulge field counterparts poses a constraint on the progenitor's mass, which is well within the range of theoretical predictions for the distribution of its stars in the $\alpha$-Fe plane.

%{\cyan [[I think we want to be more careful here.  Recall the point made by Nataf, that our metallicities are quite a bit lower than Ferraro's group.  That being the case, the age of the youngest population in Terzan~5 would be pushed a lot older, possibly matching that time of the sausage merger....]]}

%{\cyan [[I'm stuck.  The dynamical friction argument by Nate and Joel rules out anything formed in situ with a large mass.  If Terzan~5's progenitor had 10$^8$ solar masses (stellar), it must have had 100 times more DM?  According to them that would not have survived.  ]]}

\subsection{Terzan~5 as a disc-instability clump}

The discussion from the previous Sections suggests that Terzan~5 may be the remnant of a {\it minor} building block of the bulge. No simple formation scenario seems capable of explaining the properties of the bulge stellar populations, which may result from the coalescence of several systems, some of them accreted \citep[e.g.,][]{Horta_2021}, as discussed above, and others formed and evolved {\it in situ}.

\citet{Ferraro_2021} have recently found that the massive bulge globular cluster Liller~1 hosts a complex mix of stellar populations resembling that of Terzan~5.  They suggest that Liller~1, similarly to Terzan~5, may have been another contributor to the stellar mass budget of the bulge.  By assuming an initial mass of ${\rm \sim 10^9M_\odot}$ for the progenitors of both clusters, \cite{Ferraro_2021} suggest that as many as 10 such systems may have contributed to the total stellar mass of the bulge.  Considering chemistry alone, such a scenario could in fact be consistent with our results.  If one accepts that the progenitor of Terzan~5 contributed only about 1/10 of the stellar total bulge mass, a chemical composition difference between the two systems would not be at all surprising, assuming the other contributors underwent different histories of star formation to explain the resulting chemical composition differences observed.

%However, rather differently, evidence suggests an {\it in situ} origin for Liller~1 based on its $\alpha$-element abundances \citep{Horta_2020}. Further investigations into the detailed chemical compositions of Liller~1, similar to those carried out in this paper, are required to determine if the system was a \emph{major} building block, or if it too could have contributed a small fraction of the bulge stellar mass. 

\cite{Ferraro_2021} suggest that systems like Terzan~5 and Liller~1 are remnants of massive high-redshift clumps resulting from {\it in situ} MW disk gravitational instability, which migrated to the inner regions due to dynamical friction and coalesced with others into the bulge \citep{Noguchi_1998,Ceverino_2010}. 
%GCs can form from high-redshift clumps in galaxy disks whereby they form first generation stars accreted around their central regions which, after supernovae expel the remaining gas, are joined by second generation stars \citep{Bekki_2015}. 
VELA-3 cosmological simulations analysed in \citet{Mandelker_2014,Mandelker_2017} showed that a significant fraction of clumps with masses $\gtrapprox 10^{8.5}$ M$_\odot$ were long-lived and survived feedback during inward migration, allowing them to coalesce into the bulge.  On the observational side, \cite{Huertas2020} has recently  estimated the {\it stellar} masses of 3,000 clumps within 1,500 galaxies with $1 < z < 3$ in the CANDELS survey \citep{grogin2011,koekemoer2011}.  The clump stellar masses in their sample range between $\sim 10^7$ and $\sim 10^9~M_\odot$, following a power-law mass function with slope $-0.6$.  It is therefore not altogether improbable that a stellar system of that type could have been the precursor of Terzan~5, assuming that its core component, which may constitute as little as 0.1-1\% of the total initial stellar mass of the system, could live long enough to survive to this day as the bulk of the system is dissolved while migrating into the inner Galaxy.  Nonetheless, it is important to keep in mind that, according to \cite{Huertas2020}, 
%such systems exceed the star formation rate of the surrounding disc by just a factor of up to several, so that 
clump contribution to the total stellar mass of the disc is of the order of a 2--5\%, again suggesting that their impact on the mean disc chemistry should be negligible, which is in qualitative agreement with our result.

%Furthermore, \citet{Tamburello_2015} used ARGO simulations (which adopted feedback processes to stabilize the disc rather than affect individual clumps) to find that clumps are not a major contributor to the formation of galaxy bulges as is suggested here for the progenitor of Terzan~5, since the majority of those exhibiting migration to the galaxy centre had moderate masses ($< 10^8$ M$_\odot$). Their findings also suggest that, in order to prevent the clump from building mass throughout its migration, agreeing with a mass decrease of the Terzan~5 progenitor over time, it would need to have experienced sufficient feedback and avoided clump-clump mergers. In other words, the clump progenitor of Terzan~5 could have migrated inward while losing mass due to feedback processes, and have been unable to contribute significantly to the formation of the bulge if it had a mass roughly in the range $10^8 - 10^{8.5}$ M$_\odot$.

\subsection{Terzan~5 as a rejuvenated globular cluster}

An alternative scenario has been proposed by \cite{McKenzie_2018} and more recently by \cite{Bastian_2022}, according to which Terzan~5 may have been an old globular cluster whose orbit crossed paths with molecular clouds.  Such encounters could in principle lead to gas accretion and cooling, with subsequent  formation of potentially multiple new stellar generations, depending on the number of encounters and assuming that the cluster is not torn apart in the process.

Under this ``cluster rejuvenation'' scenario, one would expect that the chemical composition of the cluster's youngest stellar generations would reflect that of the Galactic disc itself.  Conversely, the chemistry of the cluster's oldest stars would correspond to that of the GC birth site at formation time, thus likely distinct from that of the local field population in the current cluster location, at the same [Fe/H], which seems to be corroborated by the fact that we found moderately metal-rich Terzan~5 stars to have lower [$\alpha$/Fe] than their bulge counterparts.  Unfortunately our sample does not have a sufficient number of stars over a wide range of [Fe/H] to afford a test of the similarities between sub-samples of Terzan~5 and their bulge field counterparts towards higher and lower metallicities. 
 
The rejuvenation scenario can, however, be tested once a larger Terzan~5 sample is obtained, so that quantitative comparisons with the field sample of detailed chemistry such as the one presented in this paper can be conducted within narrow age/metallicity ranges.

\section{Conclusions}
\label{sec:conclusions}

Terzan~5 is one of the most enigmatic stellar systems in the Milky Way.  Initially deemed to be a standard metal-rich bulge globular cluster, it was shown to host stellar populations in a range of age and metallicity \citep{Ferraro_2009,Origlia_2013,Massari_2014,Ferraro_2016} as well as the globular cluster-specific multiple-populations phenomenon \citep{Schiavon_2017b,Nataf_2019}.  Due to its complex nature, and in view of its high metallicity and broad similarity with the chemical composition of its co-local stellar populations, it has been proposed to be a remaining fragment of the building blocks of the Galactic bulge \citep[e.g.,][]{Ferraro_2009,Ferraro_2021}.

%Previous work \cite[e.g.,][]{Ferraro_2009,Massari_2014,Ferraro_2016} suggested that Terzan~5 is the remnant of a major building block of the bulge, predominantly on the basis of its metallicity and age distributions, as well as the similarity between the locus occupied by its stars in the $\alpha$-Fe plane and that of bulge field populations. 

In this paper we report a test of this hypothesis, based on a comparison between the abundance patterns of Terzan~5 and the bulge field populations.  We examine the abundance pattern of Terzan~5 based on APOGEE-2 DR17 data for 7 candidate cluster members.  The APOGEE abundances for $\alpha$-elements such as Mg, Si, Ca, and S confirm the finding by \cite{Massari_2014} that Terzan~5 and the bulge field population have a similar distribution in the $\alpha$-Fe plane, in particular regarding the metallicity of the so-called $\alpha$-knee, which has been suggested to correlate with the mass of the system \citep[e.g.,][]{Tolstoy_2009}.

We next perform a quantitative comparison between the detailed chemical compositions of Terzan~5 and the bulge field, considering the abundances of 10 elements.  By adopting a random sampling technique, we find statistically significant differences between both populations, suggesting that the Terzan~5 progenitor was unlikely to have been a {\it major} contributor to the stellar content of the bulge. 
%The above result seems to be at odds with the qualitative similarity between the distributions of the two stellar systems in the $\alpha$-Fe plane reported in previous work, particularly with regards to the metallicity of the knee \citep{Ferraro_2016}. 

We use the results of the EAGLE hydrodynamical cosmological simulations to alleviate this apparent tension by showing that the correlation between [Fe/H]$_{\rm knee}$ and M$_\star$ has a very large scatter.  We show that if indeed Terzan~5 would be the remnant of an accreted satellite, the mass of the progenitor ranges between $10^8$ and $10^{10}~{\rm M}_\odot$.  
%in excellent agreement with a previous estimate by \cite{Ferraro_2016}.  

These mass estimates are combined with the chemical composition analysis to elaborate on the origin of the Terzan~5 system.  We rule out the possibility that its progenitor could have been a massive ($\sim10^9-10^{10}{\rm M}_\odot$) satellite galaxy accreted to the Milky Way, as that would require the progenitor to be characterised by unusual abundance ratio gradients.  We argue that a relatively small progenitor mass ($\sim10^8-10^9{\rm M}_\odot$) can explain the disagreement between the chemistry of Terzan~5 and that of the bulge field.  Moreover, a relatively small progenitor mass is in qualitative agreement with scenarios proposed in the literature, according to which Terzan~5 (${\rm M\sim10^6~M_\odot}$) is the remnant of a massive stellar clump (${\rm M\sim10^8-10^9~M_\odot}$) formed in the Galactic disc at high redshift and later migrated towards the Galactic bulge while losing almost the entirety of its stellar mass, without making a major contribution to the stellar content of the inner Galaxy.

We also briefly examine the proposition by \cite{McKenzie_2018} and \cite{Bastian_2022} that Terzan~5 is instead an old globular cluster that underwent a process of rejuvenation through recent star formation due to accretion of fresh gas resulting from encounters with molecular clouds.  

While our data cannot rule out either the massive clump or the rejuvenation scenario, we conclude by proposing an observational test that may provide a decision between the scenarios discussed in this paper.  Cluster rejuvenation differs from the satellite accretion and/or massive clump scenarios in one key aspect.  Under cluster rejuvenation, young/metal-rich populations must share the chemical properties of their field counterparts, whereas the abundance ratios of the old/metal-poor Terzan~5 stars should be different, as they reflect the properties of the original formation site.  Conversely, under satellite accretion and/or massive clump formation, one would expect the chemical differences such as those identified in this paper to be present over all ages and metallicities.  An expansion of the current Terzan~5 sample of detailed elemental abundances by an order of magnitude covering the entire range of Terzan~5 metallicities should render such a test feasible.

Finally, we highlight what is perhaps the most intriguing finding in this paper, namely that Terzan~5 seems to have a {\it lower} [$\alpha$/Fe] than the field stars within the Galactic bulge.  This suggests that the Terzan~5  progenitor underwent a more prolonged star formation history, at a lower rate, than its field counterparts.  If confirmed by analysis of larger high quality samples, this result introduces an interesting and likely challenging new constraint on Terzan~5 formation scenarios.

%On the other hand, the similar [Fe/H]$_\textrm{knee}$ of Terzan~5 and the bulge indicates an undeniable similarity in their respected chemical evolution, presenting an issue that requires careful reconciliation.\\

Future investigations into the nature of complex globular cluster-like systems such as Terzan~5 (and Liller~1) would undoubtedly benefit from larger samples and detailed chemical abundances of stars belonging to those systems. Statistical comparisons with the bulge field population could then be determined with improved precision. In addition, additional insights into the nature of these systems will be gained from expanding the data to a wider range of element abundances, including diagnostics of enrichment from additional nucleosynthetic pathways not explored in this study. 

%Similar investigations into Liller~1 and other possible complex stellar systems in the inner Galaxy should be conducted, as they too may have contributed to the stellar content of the bulge.
    
%%%%%%%%%%%%%%%%%%%%%%%%%%%%%%%%%%%%%%%%%%%%%%%%%%
% \FloatBarrier
\section*{Acknowledgements}

The authors acknowledge the guidance provided by Rob Crain on the interpretation of the results from the EAGLE numerical simulations, by Dan Perley on the statistical analysis, and by Holger Baumgardt on the tidal vs Jacoby radius of Terzan~5.  Francesco Ferraro, Emanuele Dalessandro, and Nate Bastian are thanked for invaluable comments on a first draft of the manuscript.  We are also indebted to the anonymous referee for a very careful and helpful review.

J.G.F-T gratefully acknowledges the grant support provided by Proyecto Fondecyt Iniciaci\'on No. 11220340, and also from ANID Concurso de Fomento a la Vinculaci\'on Internacional para Instituciones de Investigaci\'on Regionales (Modalidad corta duraci\'on) Proyecto No. FOVI210020, and from the Joint Committee ESO-Government of Chile 2021 (ORP 023/2021). 

Funding for the Sloan Digital Sky Survey IV has been provided by the Alfred P. Sloan Foundation, the U.S. 
Department of Energy Office of Science, and the Participating Institutions. 

SDSS-IV acknowledges support and resources from the Center for High Performance Computing  at the University of Utah. The SDSS website is www.sdss.org.

SDSS-IV is managed by the Astrophysical Research Consortium for the Participating Institutions of the SDSS Collaboration including the Brazilian Participation Group, the Carnegie Institution for Science, Carnegie Mellon University, Center for Astrophysics | Harvard \& Smithsonian, the Chilean Participation Group, the French Participation Group, Instituto de Astrof\'isica de Canarias, The Johns Hopkins University, Kavli Institute for the Physics and Mathematics of the Universe (IPMU) / University of Tokyo, the Korean Participation Group, Lawrence Berkeley National Laboratory, Leibniz Institut f\"ur Astrophysik 
Potsdam (AIP),  Max-Planck-Institut f\"ur Astronomie (MPIA Heidelberg), Max-Planck-Institut f\"ur 
Astrophysik (MPA Garching), Max-Planck-Institut f\"ur Extraterrestrische Physik (MPE),
National Astronomical Observatories of China, New Mexico State University, New York University, University of 
Notre Dame, Observat\'ario Nacional / MCTI, The Ohio State University, Pennsylvania State 
University, Shanghai Astronomical Observatory, United Kingdom Participation Group, 
Universidad Nacional Aut\'onoma de M\'exico, University of Arizona, University of Colorado Boulder, 
University of Oxford, University of Portsmouth, University of Utah, University of Virginia, University
of Washington, University of Wisconsin, Vanderbilt University, and Yale University.

%%%%%%%%%%%%%%%%%%%%%%%%%%%%%%%%%%%%%%%%%%%%%%%%%%
% \FloatBarrier
\section*{Data Availability}

This work is entirely based on the 17$^{\rm th}$ data release of the SDSS-IV/APOGEE~2 survey.  All the data are publicly available at \url{https://www.sdss.org/dr17/}.



\bibliographystyle{mnras}
\bibliography{example} % if your bibtex file is called example.bib

\begin{thebibliography}{}
\makeatletter
\relax
\def\mn@urlcharsother{\let\do\@makeother \do\$\do\&\do\#\do\^\do\_\do\%\do\~}
\def\mn@doi{\begingroup\mn@urlcharsother \@ifnextchar [ {\mn@doi@}
  {\mn@doi@[]}}
\def\mn@doi@[#1]#2{\def\@tempa{#1}\ifx\@tempa\@empty \href
  {http://dx.doi.org/#2} {doi:#2}\else \href {http://dx.doi.org/#2} {#1}\fi
  \endgroup}
\def\mn@eprint#1#2{\mn@eprint@#1:#2::\@nil}
\def\mn@eprint@arXiv#1{\href {http://arxiv.org/abs/#1} {{\tt arXiv:#1}}}
\def\mn@eprint@dblp#1{\href {http://dblp.uni-trier.de/rec/bibtex/#1.xml}
  {dblp:#1}}
\def\mn@eprint@#1:#2:#3:#4\@nil{\def\@tempa {#1}\def\@tempb {#2}\def\@tempc
  {#3}\ifx \@tempc \@empty \let \@tempc \@tempb \let \@tempb \@tempa \fi \ifx
  \@tempb \@empty \def\@tempb {arXiv}\fi \@ifundefined
  {mn@eprint@\@tempb}{\@tempb:\@tempc}{\expandafter \expandafter \csname
  mn@eprint@\@tempb\endcsname \expandafter{\@tempc}}}

\bibitem[\protect\citeauthoryear{{Andrews}, {Weinberg}, {Sch{\"o}nrich}  \&
  {Johnson}}{{Andrews} et~al.}{2017}]{Andrews2017}
{Andrews} B.~H.,  {Weinberg} D.~H.,  {Sch{\"o}nrich} R.,   {Johnson} J.~A.,
  2017, \mn@doi [\apj] {10.3847/1538-4357/835/2/224}, \href
  {https://ui.adsabs.harvard.edu/abs/2017ApJ...835..224A} {835, 224}

\bibitem[\protect\citeauthoryear{{Barbuy}, {Bica}  \& {Ortolani}}{{Barbuy}
  et~al.}{1998}]{Barbuy_1998}
{Barbuy} B.,  {Bica} E.,   {Ortolani} S.,  1998, \aap, \href
  {https://ui.adsabs.harvard.edu/abs/1998A&A...333..117B} {333, 117}

\bibitem[\protect\citeauthoryear{{Barbuy}, {Chiappini}  \& {Gerhard}}{{Barbuy}
  et~al.}{2018}]{Barbuy_2018}
{Barbuy} B.,  {Chiappini} C.,   {Gerhard} O.,  2018, \mn@doi [\araa]
  {10.1146/annurev-astro-081817-051826}, \href
  {https://ui.adsabs.harvard.edu/abs/2018ARA&A..56..223B} {56, 223}

\bibitem[\protect\citeauthoryear{{Bastian} \& {Lardo}}{{Bastian} \&
  {Lardo}}{2018}]{BastianLardo_2018}
{Bastian} N.,  {Lardo} C.,  2018, \mn@doi [\araa]
  {10.1146/annurev-astro-081817-051839}, \href
  {https://ui.adsabs.harvard.edu/abs/2018ARA&A..56...83B} {56, 83}

\bibitem[\protect\citeauthoryear{{Bastian} \& {Pfeffer}}{{Bastian} \&
  {Pfeffer}}{2022}]{Bastian_2022}
{Bastian} N.,  {Pfeffer} J.,  2022, \mn@doi [\mnras] {10.1093/mnras/stab3081},
  \href {https://ui.adsabs.harvard.edu/abs/2022MNRAS.509..614B} {509, 614}

\bibitem[\protect\citeauthoryear{Bastian et~al.,}{Bastian
  et~al.}{2020}]{Bastian_2020}
Bastian N.,  et~al., 2020, \mn@doi [Monthly Notices of the Royal Astronomical
  Society] {10.1093/mnras/staa716}, 494, 332–337

\bibitem[\protect\citeauthoryear{Baumgardt, Sollima, Hilker, Bellini  \&
  Vasiliev}{Baumgardt et~al.}{2021}]{Baumgardt_2021}
Baumgardt H.,  Sollima A.,  Hilker M.,  Bellini A.,   Vasiliev E.,  2021,
  Fundamental parameters of Galactic globular clusters, \url
  {https://people.smp.uq.edu.au/HolgerBaumgardt/globular/}

\bibitem[\protect\citeauthoryear{Bellini, Piotto, Bedin, King, Anderson, Milone
   \& Momany}{Bellini et~al.}{2009}]{Bellini_2009}
Bellini A.,  Piotto G.,  Bedin L.~R.,  King I.~R.,  Anderson J.,  Milone A.~P.,
    Momany Y.,  2009, \mn@doi [Astronomy & Astrophysics]
  {10.1051/0004-6361/200912757}, 507, 1393–1408

\bibitem[\protect\citeauthoryear{Bellini, Bedin, Piotto, Milone, Marino  \&
  Villanova}{Bellini et~al.}{2010}]{Bellini_2010}
Bellini A.,  Bedin L.~R.,  Piotto G.,  Milone A.~P.,  Marino A.~F.,   Villanova
  S.,  2010, \mn@doi [The Astronomical Journal] {10.1088/0004-6256/140/2/631},
  140, 631–641

\bibitem[\protect\citeauthoryear{Bellini, Anderson, Salaris, Cassisi, Bedin,
  Piotto  \& Bergeron}{Bellini et~al.}{2013}]{Bellini_2013}
Bellini A.,  Anderson J.,  Salaris M.,  Cassisi S.,  Bedin L.~R.,  Piotto G.,
  Bergeron P.,  2013, \mn@doi [The Astrophysical Journal]
  {10.1088/2041-8205/769/2/l32}, 769, L32

\bibitem[\protect\citeauthoryear{{Belokurov}, {Erkal}, {Evans}, {Koposov}  \&
  {Deason}}{{Belokurov} et~al.}{2018}]{Belokurov_2018}
{Belokurov} V.,  {Erkal} D.,  {Evans} N.~W.,  {Koposov} S.~E.,   {Deason}
  A.~J.,  2018, \mn@doi [\mnras] {10.1093/mnras/sty982}, \href
  {https://ui.adsabs.harvard.edu/abs/2018MNRAS.478..611B} {478, 611}

\bibitem[\protect\citeauthoryear{{Bensby}, {Feltzing}, {Yee}, {Johnson},
  {Gould}, {Asplund}, {Mel{\'e}ndez}  \& {Lucatello}}{{Bensby}
  et~al.}{2020}]{Bensby_2020}
{Bensby} T.,  {Feltzing} S.,  {Yee} J.~C.,  {Johnson} J.~A.,  {Gould} A.,
  {Asplund} M.,  {Mel{\'e}ndez} J.,   {Lucatello} S.,  2020, \mn@doi [\aap]
  {10.1051/0004-6361/201937401}, \href
  {https://ui.adsabs.harvard.edu/abs/2020A&A...634A.130B} {634, A130}

\bibitem[\protect\citeauthoryear{{Blanton} et~al.,}{{Blanton}
  et~al.}{2017}]{Blanton_2017}
{Blanton} M.~R.,  et~al., 2017, \mn@doi [\aj] {10.3847/1538-3881/aa7567}, \href
  {https://ui.adsabs.harvard.edu/abs/2017AJ....154...28B} {154, 28}

\bibitem[\protect\citeauthoryear{{Bowen} \& {Vaughan}}{{Bowen} \&
  {Vaughan}}{1973}]{Bowen_1973}
{Bowen} I.~S.,  {Vaughan} A.~H. J.,  1973, \mn@doi [\ao]
  {10.1364/AO.12.001430}, \href
  {https://ui.adsabs.harvard.edu/abs/1973ApOpt..12.1430B} {12, 1430}

\bibitem[\protect\citeauthoryear{Carretta et~al.,}{Carretta
  et~al.}{2010}]{Carretta_2010}
Carretta E.,  et~al., 2010, \mn@doi [The Astrophysical Journal]
  {10.1088/2041-8205/714/1/l7}, 714, L7–L11

\bibitem[\protect\citeauthoryear{{Ceverino}, {Dekel}  \& {Bournaud}}{{Ceverino}
  et~al.}{2010}]{Ceverino_2010}
{Ceverino} D.,  {Dekel} A.,   {Bournaud} F.,  2010, \mn@doi [\mnras]
  {10.1111/j.1365-2966.2010.16433.x}, \href
  {https://ui.adsabs.harvard.edu/abs/2010MNRAS.404.2151C} {404, 2151}

\bibitem[\protect\citeauthoryear{Cleveland}{Cleveland}{1979}]{Cleveland_1979}
Cleveland W.~S.,  1979, \mn@doi [Journal of the American Statistical
  Association] {10.1080/01621459.1979.10481038}, 74, 829

\bibitem[\protect\citeauthoryear{{Crain} et~al.,}{{Crain}
  et~al.}{2015}]{crain_2015}
{Crain} R.~A.,  et~al., 2015, \mn@doi [\mnras] {10.1093/mnras/stv725}, \href
  {https://ui.adsabs.harvard.edu/abs/2015MNRAS.450.1937C} {450, 1937}

\bibitem[\protect\citeauthoryear{{Eilers}, {Hogg}, {Rix}, {Ness},
  {Price-Whelan}, {Meszaros}  \& {Nitschelm}}{{Eilers}
  et~al.}{2021}]{Eilers2021}
{Eilers} A.-C.,  {Hogg} D.~W.,  {Rix} H.-W.,  {Ness} M.~K.,  {Price-Whelan}
  A.~M.,  {Meszaros} S.,   {Nitschelm} C.,  2021, arXiv e-prints, \href
  {https://ui.adsabs.harvard.edu/abs/2021arXiv211203295E} {p. arXiv:2112.03295}

\bibitem[\protect\citeauthoryear{Fernández-Trincado
  et~al.,}{Fernández-Trincado et~al.}{2021}]{Trincado_2021}
Fernández-Trincado J.~G.,  et~al., 2021, \mn@doi [Astronomy & Astrophysics]
  {10.1051/0004-6361/202140306}, 648, A70

\bibitem[\protect\citeauthoryear{Ferraro, Sollima, Pancino, Bellazzini,
  Straniero, Origlia  \& Cool}{Ferraro et~al.}{2004}]{Ferraro_2004}
Ferraro F.~R.,  Sollima A.,  Pancino E.,  Bellazzini M.,  Straniero O.,
  Origlia L.,   Cool A.~M.,  2004, \mn@doi [The Astrophysical Journal]
  {10.1086/383149}, 603, L81–L84

\bibitem[\protect\citeauthoryear{Ferraro, Sollima, Rood, Origlia, Pancino  \&
  Bellazzini}{Ferraro et~al.}{2006}]{Ferraro_2006}
Ferraro F.~R.,  Sollima A.,  Rood R.~T.,  Origlia L.,  Pancino E.,   Bellazzini
  M.,  2006, \mn@doi [The Astrophysical Journal] {10.1086/498735}, 638,
  433–439

\bibitem[\protect\citeauthoryear{Ferraro et~al.,}{Ferraro
  et~al.}{2009}]{Ferraro_2009}
Ferraro F.~R.,  et~al., 2009, \mn@doi [Nature] {10.1038/nature08581}, 462,
  483–486

\bibitem[\protect\citeauthoryear{Ferraro, Massari, Dalessandro, Lanzoni,
  Origlia, Rich  \& Mucciarelli}{Ferraro et~al.}{2016}]{Ferraro_2016}
Ferraro F.~R.,  Massari D.,  Dalessandro E.,  Lanzoni B.,  Origlia L.,  Rich
  R.~M.,   Mucciarelli A.,  2016, \mn@doi [The Astrophysical Journal]
  {10.3847/0004-637x/828/2/75}, 828, 75

\bibitem[\protect\citeauthoryear{{Ferraro} et~al.,}{{Ferraro}
  et~al.}{2021}]{Ferraro_2021}
{Ferraro} F.~R.,  et~al., 2021, \mn@doi [Nature Astronomy]
  {10.1038/s41550-020-01267-y}, \href
  {https://ui.adsabs.harvard.edu/abs/2021NatAs...5..311F} {5, 311}

\bibitem[\protect\citeauthoryear{{Gaia Collaboration} et~al.,}{{Gaia
  Collaboration} et~al.}{2016}]{Gaia_2016}
{Gaia Collaboration} et~al., 2016, \mn@doi [Astronomy & Astrophysics]
  {10.1051/0004-6361/201629272}, 595, A1

\bibitem[\protect\citeauthoryear{{Gaia Collaboration} et~al.,}{{Gaia
  Collaboration} et~al.}{2017}]{Gaia_2018b}
{Gaia Collaboration} et~al., 2017, \mn@doi [\aap]
  {10.1051/0004-6361/201629925}, \href
  {https://ui.adsabs.harvard.edu/abs/2017A&A...605A..79G} {605, A79}

\bibitem[\protect\citeauthoryear{{Gaia Collaboration} et~al.,}{{Gaia
  Collaboration} et~al.}{2021a}]{Gaia_2021_sum}
{Gaia Collaboration} et~al., 2021a, \mn@doi [\aap]
  {10.1051/0004-6361/202039657}, \href
  {https://ui.adsabs.harvard.edu/abs/2021A&A...649A...1G} {649, A1}

\bibitem[\protect\citeauthoryear{{Gaia Collaboration} et~al.,}{{Gaia
  Collaboration} et~al.}{2021b}]{Gaia_2021_phot}
{Gaia Collaboration} et~al., 2021b, \mn@doi [\aap]
  {10.1051/0004-6361/202039709}, \href
  {https://ui.adsabs.harvard.edu/abs/2021A&A...649A...2L} {649, A2}

\bibitem[\protect\citeauthoryear{{Garc{\'\i}a P{\'e}rez} et~al.,}{{Garc{\'\i}a
  P{\'e}rez} et~al.}{2016}]{Perez_2016}
{Garc{\'\i}a P{\'e}rez} A.~E.,  et~al., 2016, \mn@doi [\aj]
  {10.3847/0004-6256/151/6/144}, \href
  {https://ui.adsabs.harvard.edu/abs/2016AJ....151..144G} {151, 144}

\bibitem[\protect\citeauthoryear{{Geisler} et~al.,}{{Geisler}
  et~al.}{2021}]{Geisler_2021}
{Geisler} D.,  et~al., 2021, \mn@doi [\aap] {10.1051/0004-6361/202140436},
  \href {https://ui.adsabs.harvard.edu/abs/2021A&A...652A.157G} {652, A157}

\bibitem[\protect\citeauthoryear{{Greggio} \& {Renzini}}{{Greggio} \&
  {Renzini}}{1983}]{Greggio_1983}
{Greggio} L.,  {Renzini} A.,  1983, \aap, \href
  {https://ui.adsabs.harvard.edu/abs/1983A&A...118..217G} {118, 217}

\bibitem[\protect\citeauthoryear{{Grogin} et~al.,}{{Grogin}
  et~al.}{2011}]{grogin2011}
{Grogin} N.~A.,  et~al., 2011, \mn@doi [\apjs] {10.1088/0067-0049/197/2/35},
  \href {https://ui.adsabs.harvard.edu/abs/2011ApJS..197...35G} {197, 35}

\bibitem[\protect\citeauthoryear{{Gunn} et~al.,}{{Gunn}
  et~al.}{2006}]{Gunn_2006}
{Gunn} J.~E.,  et~al., 2006, \mn@doi [\aj] {10.1086/500975}, \href
  {https://ui.adsabs.harvard.edu/abs/2006AJ....131.2332G} {131, 2332}

\bibitem[\protect\citeauthoryear{Harris}{Harris}{2010}]{harris2010new}
Harris W.~E.,  2010, A New Catalog of Globular Clusters in the Milky Way
  (\mn@eprint {arXiv} {1012.3224})

\bibitem[\protect\citeauthoryear{{Hayes} et~al.,}{{Hayes}
  et~al.}{2020}]{Hayes_2020}
{Hayes} C.~R.,  et~al., 2020, \mn@doi [\apj] {10.3847/1538-4357/ab62ad}, \href
  {https://ui.adsabs.harvard.edu/abs/2020ApJ...889...63H} {889, 63}

\bibitem[\protect\citeauthoryear{{Haywood}, {Di Matteo}, {Lehnert}, {Snaith},
  {Khoperskov}  \& {G{\'o}mez}}{{Haywood} et~al.}{2018}]{Haywood_2018}
{Haywood} M.,  {Di Matteo} P.,  {Lehnert} M.~D.,  {Snaith} O.,  {Khoperskov}
  S.,   {G{\'o}mez} A.,  2018, \mn@doi [\apj] {10.3847/1538-4357/aad235}, \href
  {https://ui.adsabs.harvard.edu/abs/2018ApJ...863..113H} {863, 113}

\bibitem[\protect\citeauthoryear{{Helmi}, {Babusiaux}, {Koppelman}, {Massari},
  {Veljanoski}  \& {Brown}}{{Helmi} et~al.}{2018}]{Helmi_2018}
{Helmi} A.,  {Babusiaux} C.,  {Koppelman} H.~H.,  {Massari} D.,  {Veljanoski}
  J.,   {Brown} A. G.~A.,  2018, \mn@doi [\nat] {10.1038/s41586-018-0625-x},
  \href {https://ui.adsabs.harvard.edu/abs/2018Natur.563...85H} {563, 85}

\bibitem[\protect\citeauthoryear{{Hendricks}, {Koch}, {Lanfranchi}, {Boeche},
  {Walker}, {Johnson}, {Pe{\~n}arrubia}  \& {Gilmore}}{{Hendricks}
  et~al.}{2014}]{Hendricks_2014}
{Hendricks} B.,  {Koch} A.,  {Lanfranchi} G.~A.,  {Boeche} C.,  {Walker} M.,
  {Johnson} C.~I.,  {Pe{\~n}arrubia} J.,   {Gilmore} G.,  2014, \mn@doi [\apj]
  {10.1088/0004-637X/785/2/102}, \href
  {https://ui.adsabs.harvard.edu/abs/2014ApJ...785..102H} {785, 102}

\bibitem[\protect\citeauthoryear{{Holtzman} et~al.,}{{Holtzman}
  et~al.}{2015}]{Holtzman_2015}
{Holtzman} J.~A.,  et~al., 2015, \mn@doi [\aj] {10.1088/0004-6256/150/5/148},
  \href {https://ui.adsabs.harvard.edu/abs/2015AJ....150..148H} {150, 148}

\bibitem[\protect\citeauthoryear{{Holtzman} et~al.,}{{Holtzman}
  et~al.}{2018}]{Holtzman2018}
{Holtzman} J.~A.,  et~al., 2018, \mn@doi [\aj] {10.3847/1538-3881/aad4f9},
  \href {https://ui.adsabs.harvard.edu/abs/2018AJ....156..125H} {156, 125}

\bibitem[\protect\citeauthoryear{{Horta} et~al.,}{{Horta}
  et~al.}{2021}]{Horta_2021}
{Horta} D.,  et~al., 2021, \mn@doi [\mnras] {10.1093/mnras/staa2987}, \href
  {https://ui.adsabs.harvard.edu/abs/2021MNRAS.500.1385H} {500, 1385}

\bibitem[\protect\citeauthoryear{{Huertas-Company} et~al.,}{{Huertas-Company}
  et~al.}{2020}]{Huertas2020}
{Huertas-Company} M.,  et~al., 2020, \mn@doi [\mnras] {10.1093/mnras/staa2777},
  \href {https://ui.adsabs.harvard.edu/abs/2020MNRAS.499..814H} {499, 814}

\bibitem[\protect\citeauthoryear{{Ibata}, {Gilmore}  \& {Irwin}}{{Ibata}
  et~al.}{1994}]{Ibata_1994}
{Ibata} R.~A.,  {Gilmore} G.,   {Irwin} M.~J.,  1994, \mn@doi [\nat]
  {10.1038/370194a0}, \href
  {https://ui.adsabs.harvard.edu/abs/1994Natur.370..194I} {370, 194}

\bibitem[\protect\citeauthoryear{Immeli, Samland, Westera  \& Gerhard}{Immeli
  et~al.}{2004}]{Immeli_2004}
Immeli A.,  Samland M.,  Westera P.,   Gerhard O.,  2004, \mn@doi [The
  Astrophysical Journal] {10.1086/422179}, 611, 20–25

\bibitem[\protect\citeauthoryear{{Johnson}, {Caldwell}, {Rich}, {Mateo},
  {Bailey}, {Clarkson}, {Olszewski}  \& {Walker}}{{Johnson}
  et~al.}{2017}]{Johnson_2017}
{Johnson} C.~I.,  {Caldwell} N.,  {Rich} R.~M.,  {Mateo} M.,  {Bailey} John~I.
  I.,  {Clarkson} W.~I.,  {Olszewski} E.~W.,   {Walker} M.~G.,  2017, \mn@doi
  [\apj] {10.3847/1538-4357/836/2/168}, \href
  {https://ui.adsabs.harvard.edu/abs/2017ApJ...836..168J} {836, 168}

\bibitem[\protect\citeauthoryear{{J{\"o}nsson} et~al.,}{{J{\"o}nsson}
  et~al.}{2020}]{Jonsson2020}
{J{\"o}nsson} H.,  et~al., 2020, \mn@doi [\aj] {10.3847/1538-3881/aba592},
  \href {https://ui.adsabs.harvard.edu/abs/2020AJ....160..120J} {160, 120}

\bibitem[\protect\citeauthoryear{{King}}{{King}}{1966}]{King_1966}
{King} I.~R.,  1966, \mn@doi [\aj] {10.1086/109918}, \href
  {https://ui.adsabs.harvard.edu/abs/1966AJ.....71..276K} {71, 276}

\bibitem[\protect\citeauthoryear{{Koekemoer} et~al.,}{{Koekemoer}
  et~al.}{2011}]{koekemoer2011}
{Koekemoer} A.~M.,  et~al., 2011, \mn@doi [\apjs] {10.1088/0067-0049/197/2/36},
  \href {https://ui.adsabs.harvard.edu/abs/2011ApJS..197...36K} {197, 36}

\bibitem[\protect\citeauthoryear{Kormendy \& Kennicutt}{Kormendy \&
  Kennicutt}{2004}]{Kormendy_2004}
Kormendy J.,  Kennicutt R.~C.,  2004, \mn@doi [Annual Review of Astronomy and
  Astrophysics] {10.1146/annurev.astro.42.053102.134024}, 42, 603–683

\bibitem[\protect\citeauthoryear{Lanzoni et~al.,}{Lanzoni
  et~al.}{2010}]{Lanzoni_2010}
Lanzoni B.,  et~al., 2010, \mn@doi [The Astrophysical Journal]
  {10.1088/0004-637x/717/2/653}, 717, 653–657

\bibitem[\protect\citeauthoryear{Lee, Joo, Sohn, Rey, Lee  \& Walker}{Lee
  et~al.}{1999}]{Lee_1999}
Lee Y.-W.,  Joo J.-M.,  Sohn Y.-J.,  Rey S.-C.,  Lee H.-c.,   Walker A.~R.,
  1999, \mn@doi [Nature] {10.1038/46985}, 402, 55–57

\bibitem[\protect\citeauthoryear{{Leung} \& {Bovy}}{{Leung} \&
  {Bovy}}{2019a}]{LeungBovy_2019a}
{Leung} H.~W.,  {Bovy} J.,  2019a, \mn@doi [\mnras] {10.1093/mnras/sty3217},
  \href {https://ui.adsabs.harvard.edu/abs/2019MNRAS.483.3255L} {483, 3255}

\bibitem[\protect\citeauthoryear{{Leung} \& {Bovy}}{{Leung} \&
  {Bovy}}{2019b}]{LeungBovy_2019b}
{Leung} H.~W.,  {Bovy} J.,  2019b, \mn@doi [\mnras] {10.1093/mnras/stz2245},
  \href {https://ui.adsabs.harvard.edu/abs/2019MNRAS.489.2079L} {489, 2079}

\bibitem[\protect\citeauthoryear{{Mackereth} et~al.,}{{Mackereth}
  et~al.}{2019}]{Mackereth_2019}
{Mackereth} J.~T.,  et~al., 2019, \mn@doi [\mnras] {10.1093/mnras/sty2955},
  \href {https://ui.adsabs.harvard.edu/abs/2019MNRAS.482.3426M} {482, 3426}

\bibitem[\protect\citeauthoryear{{Majewski} et~al.,}{{Majewski}
  et~al.}{2017}]{Majewski_2017}
{Majewski} S.~R.,  et~al., 2017, \mn@doi [\aj] {10.3847/1538-3881/aa784d},
  \href {https://ui.adsabs.harvard.edu/abs/2017AJ....154...94M} {154, 94}

\bibitem[\protect\citeauthoryear{{Mandelker}, {Dekel}, {Ceverino}, {Tweed},
  {Moody}  \& {Primack}}{{Mandelker} et~al.}{2014}]{Mandelker_2014}
{Mandelker} N.,  {Dekel} A.,  {Ceverino} D.,  {Tweed} D.,  {Moody} C.~E.,
  {Primack} J.,  2014, \mn@doi [\mnras] {10.1093/mnras/stu1340}, \href
  {https://ui.adsabs.harvard.edu/abs/2014MNRAS.443.3675M} {443, 3675}

\bibitem[\protect\citeauthoryear{{Mandelker}, {Dekel}, {Ceverino}, {DeGraf},
  {Guo}  \& {Primack}}{{Mandelker} et~al.}{2017}]{Mandelker_2017}
{Mandelker} N.,  {Dekel} A.,  {Ceverino} D.,  {DeGraf} C.,  {Guo} Y.,
  {Primack} J.,  2017, \mn@doi [\mnras] {10.1093/mnras/stw2358}, \href
  {https://ui.adsabs.harvard.edu/abs/2017MNRAS.464..635M} {464, 635}

\bibitem[\protect\citeauthoryear{{Mannucci}, {Della Valle}, {Panagia},
  {Cappellaro}, {Cresci}, {Maiolino}, {Petrosian}  \& {Turatto}}{{Mannucci}
  et~al.}{2005}]{Mannucci_2005}
{Mannucci} F.,  {Della Valle} M.,  {Panagia} N.,  {Cappellaro} E.,  {Cresci}
  G.,  {Maiolino} R.,  {Petrosian} A.,   {Turatto} M.,  2005, \mn@doi [\aap]
  {10.1051/0004-6361:20041411}, \href
  {https://ui.adsabs.harvard.edu/abs/2005A&A...433..807M} {433, 807}

\bibitem[\protect\citeauthoryear{{Maoz}, {Mannucci}, {Li}, {Filippenko}, {Della
  Valle}  \& {Panagia}}{{Maoz} et~al.}{2011}]{Maoz_2011}
{Maoz} D.,  {Mannucci} F.,  {Li} W.,  {Filippenko} A.~V.,  {Della Valle} M.,
  {Panagia} N.,  2011, \mn@doi [\mnras] {10.1111/j.1365-2966.2010.16808.x},
  \href {https://ui.adsabs.harvard.edu/abs/2011MNRAS.412.1508M} {412, 1508}

\bibitem[\protect\citeauthoryear{Marino, Milone, Piotto, Villanova, Bedin,
  Bellini  \& Renzini}{Marino et~al.}{2009}]{Marino_2009}
Marino A.~F.,  Milone A.~P.,  Piotto G.,  Villanova S.,  Bedin L.~R.,  Bellini
  A.,   Renzini A.,  2009, \mn@doi [Astronomy & Astrophysics]
  {10.1051/0004-6361/200911827}, 505, 1099–1113

\bibitem[\protect\citeauthoryear{Marino et~al.,}{Marino
  et~al.}{2011}]{Marino_2011}
Marino A.~F.,  et~al., 2011, \mn@doi [Astronomy & Astrophysics]
  {10.1051/0004-6361/201116546}, 532, A8

\bibitem[\protect\citeauthoryear{Marino et~al.,}{Marino
  et~al.}{2012}]{Marino_2012}
Marino A.~F.,  et~al., 2012, \mn@doi [Astronomy & Astrophysics]
  {10.1051/0004-6361/201118381}, 541, A15

\bibitem[\protect\citeauthoryear{Massari, Ferraro, Mucciarelli, Origlia,
  Dalessandro  \& Lanzoni}{Massari et~al.}{2014}]{Massari_2014}
Massari D.,  Ferraro F.~R.,  Mucciarelli A.,  Origlia L.,  Dalessandro E.,
  Lanzoni B.,  2014, Terzan 5: a pristine fragment of the Galactic Bulge?
  (\mn@eprint {arXiv} {1401.3962})

\bibitem[\protect\citeauthoryear{{Matteucci} \& {Greggio}}{{Matteucci} \&
  {Greggio}}{1986}]{Matteucci_1986}
{Matteucci} F.,  {Greggio} L.,  1986, \aap, \href
  {https://ui.adsabs.harvard.edu/abs/1986A&A...154..279M} {154, 279}

\bibitem[\protect\citeauthoryear{{McConnachie}}{{McConnachie}}{2012}]{mcconnachie2012}
{McConnachie} A.~W.,  2012, \mn@doi [\aj] {10.1088/0004-6256/144/1/4}, \href
  {https://ui.adsabs.harvard.edu/abs/2012AJ....144....4M} {144, 4}

\bibitem[\protect\citeauthoryear{McKenzie \& Bekki}{McKenzie \&
  Bekki}{2018}]{McKenzie_2018}
McKenzie M.,  Bekki K.,  2018, \mn@doi [Monthly Notices of the Royal
  Astronomical Society] {10.1093/mnras/sty1557}, 479, 3126–3141

\bibitem[\protect\citeauthoryear{{McWilliam}}{{McWilliam}}{1997}]{McWilliam1997}
{McWilliam} A.,  1997, \mn@doi [\araa] {10.1146/annurev.astro.35.1.503}, \href
  {https://ui.adsabs.harvard.edu/abs/1997ARA&A..35..503M} {35, 503}

\bibitem[\protect\citeauthoryear{{Minniti}}{{Minniti}}{1995}]{Minniti_1995}
{Minniti} D.,  1995, \mn@doi [\aj] {10.1086/117393}, \href
  {https://ui.adsabs.harvard.edu/abs/1995AJ....109.1663M} {109, 1663}

\bibitem[\protect\citeauthoryear{{Minniti} et~al.,}{{Minniti}
  et~al.}{2010}]{Minniti_2010}
{Minniti} D.,  et~al., 2010, \mn@doi [\na] {10.1016/j.newast.2009.12.002},
  \href {https://ui.adsabs.harvard.edu/abs/2010NewA...15..433M} {15, 433}

\bibitem[\protect\citeauthoryear{{Nataf}}{{Nataf}}{2017}]{Nataf_2017}
{Nataf} D.~M.,  2017, \mn@doi [\pasa] {10.1017/pasa.2017.32}, \href
  {https://ui.adsabs.harvard.edu/abs/2017PASA...34...41N} {34, e041}

\bibitem[\protect\citeauthoryear{Nataf et~al.,}{Nataf
  et~al.}{2019}]{Nataf_2019}
Nataf et~al., 2019, \mn@doi [The Astronomical Journal]
  {10.3847/1538-3881/ab1a27}, 158, 14

\bibitem[\protect\citeauthoryear{{Nidever} et~al.,}{{Nidever}
  et~al.}{2020}]{Nidever_2020}
{Nidever} D.~L.,  et~al., 2020, \mn@doi [\apj] {10.3847/1538-4357/ab7305},
  \href {https://ui.adsabs.harvard.edu/abs/2020ApJ...895...88N} {895, 88}

\bibitem[\protect\citeauthoryear{{Noguchi}}{{Noguchi}}{1998}]{Noguchi_1998}
{Noguchi} M.,  1998, \mn@doi [\nat] {10.1038/32596}, \href
  {https://ui.adsabs.harvard.edu/abs/1998Natur.392..253N} {392, 253}

\bibitem[\protect\citeauthoryear{{Origlia} et~al.,}{{Origlia}
  et~al.}{2011a}]{Origlia_2010}
{Origlia} L.,  et~al., 2011a, \mn@doi [\apjl] {10.1088/2041-8205/726/2/L20},
  \href {https://ui.adsabs.harvard.edu/abs/2011ApJ...726L..20O} {726, L20}

\bibitem[\protect\citeauthoryear{{Origlia} et~al.,}{{Origlia}
  et~al.}{2011b}]{Origlia2011}
{Origlia} L.,  et~al., 2011b, \mn@doi [\apjl] {10.1088/2041-8205/726/2/L20},
  \href {https://ui.adsabs.harvard.edu/abs/2011ApJ...726L..20O} {726, L20}

\bibitem[\protect\citeauthoryear{{Origlia}, {Massari}, {Rich}, {Mucciarelli},
  {Ferraro}, {Dalessandro}  \& {Lanzoni}}{{Origlia}
  et~al.}{2013}]{Origlia_2013}
{Origlia} L.,  {Massari} D.,  {Rich} R.~M.,  {Mucciarelli} A.,  {Ferraro}
  F.~R.,  {Dalessandro} E.,   {Lanzoni} B.,  2013, \mn@doi [\apjl]
  {10.1088/2041-8205/779/1/L5}, \href
  {https://ui.adsabs.harvard.edu/abs/2013ApJ...779L...5O} {779, L5}

\bibitem[\protect\citeauthoryear{Pancino, Ferraro, Bellazzini, Piotto  \&
  Zoccali}{Pancino et~al.}{2000}]{Pancino_2000}
Pancino E.,  Ferraro F.~R.,  Bellazzini M.,  Piotto G.,   Zoccali M.,  2000,
  \mn@doi [The Astrophysical Journal] {10.1086/312658}, 534, L83–L87

\bibitem[\protect\citeauthoryear{Pfeffer, Lardo, Bastian, Saracino  \&
  Kamann}{Pfeffer et~al.}{2020}]{Pfeffer_2020}
Pfeffer J.,  Lardo C.,  Bastian N.,  Saracino S.,   Kamann S.,  2020, \mn@doi
  [Monthly Notices of the Royal Astronomical Society] {10.1093/mnras/staa3407},
  500, 2514–2524

\bibitem[\protect\citeauthoryear{{Pfeffer}, {Lardo}, {Bastian}, {Saracino}  \&
  {Kamann}}{{Pfeffer} et~al.}{2021}]{Pfeffer_2021}
{Pfeffer} J.,  {Lardo} C.,  {Bastian} N.,  {Saracino} S.,   {Kamann} S.,  2021,
  \mn@doi [\mnras] {10.1093/mnras/staa3407}, \href
  {https://ui.adsabs.harvard.edu/abs/2021MNRAS.500.2514P} {500, 2514}

\bibitem[\protect\citeauthoryear{Prager, Ransom, Freire, Hessels, Stairs, Arras
   \& Cadelano}{Prager et~al.}{2017}]{Prager_2017}
Prager B.~J.,  Ransom S.~M.,  Freire P. C.~C.,  Hessels J. W.~T.,  Stairs
  I.~H.,  Arras P.,   Cadelano M.,  2017, \mn@doi [The Astrophysical Journal]
  {10.3847/1538-4357/aa7ed7}, 845, 148

\bibitem[\protect\citeauthoryear{Renzini}{Renzini}{2008}]{Renzini_2008}
Renzini A.,  2008, \mn@doi [Monthly Notices of the Royal Astronomical Society]
  {10.1111/j.1365-2966.2008.13892.x}, 391, 354–362

\bibitem[\protect\citeauthoryear{{Renzini} et~al.,}{{Renzini}
  et~al.}{2015}]{Renzini_2015}
{Renzini} A.,  et~al., 2015, \mn@doi [\mnras] {10.1093/mnras/stv2268}, \href
  {https://ui.adsabs.harvard.edu/abs/2015MNRAS.454.4197R} {454, 4197}

\bibitem[\protect\citeauthoryear{{Schaye} et~al.,}{{Schaye}
  et~al.}{2015}]{schaye_2015}
{Schaye} J.,  et~al., 2015, \mn@doi [\mnras] {10.1093/mnras/stu2058}, \href
  {https://ui.adsabs.harvard.edu/abs/2015MNRAS.446..521S} {446, 521}

\bibitem[\protect\citeauthoryear{{Schiavon} et~al.,}{{Schiavon}
  et~al.}{2017}]{Schiavon_2017b}
{Schiavon} R.~P.,  et~al., 2017, \mn@doi [\mnras] {10.1093/mnras/stw3093},
  \href {https://ui.adsabs.harvard.edu/abs/2017MNRAS.466.1010S} {466, 1010}

\bibitem[\protect\citeauthoryear{Shen, Rich, Kormendy, Howard, De~Propris  \&
  Kunder}{Shen et~al.}{2010}]{Shen_2010}
Shen J.,  Rich R.~M.,  Kormendy J.,  Howard C.~D.,  De~Propris R.,   Kunder A.,
   2010, \mn@doi [The Astrophysical Journal] {10.1088/2041-8205/720/1/l72},
  720, L72–L76

\bibitem[\protect\citeauthoryear{{Tinsley}}{{Tinsley}}{1979}]{Tinsley_1979}
{Tinsley} B.~M.,  1979, \mn@doi [\apj] {10.1086/157039}, \href
  {https://ui.adsabs.harvard.edu/abs/1979ApJ...229.1046T} {229, 1046}

\bibitem[\protect\citeauthoryear{{Tolstoy}, {Hill}  \& {Tosi}}{{Tolstoy}
  et~al.}{2009}]{Tolstoy_2009}
{Tolstoy} E.,  {Hill} V.,   {Tosi} M.,  2009, \mn@doi [\araa]
  {10.1146/annurev-astro-082708-101650}, \href
  {https://ui.adsabs.harvard.edu/abs/2009ARA&A..47..371T} {47, 371}

\bibitem[\protect\citeauthoryear{Utkin \& Dambis}{Utkin \&
  Dambis}{2020}]{Utkin_2020}
Utkin N.~D.,  Dambis A.~K.,  2020, \mn@doi [Monthly Notices of the Royal
  Astronomical Society] {10.1093/mnras/staa2819}, 499, 1058–1071

\bibitem[\protect\citeauthoryear{Valenti et~al.,}{Valenti
  et~al.}{2016}]{Valenti_2016}
Valenti E.,  et~al., 2016, \mn@doi [Astronomy & Astrophysics]
  {10.1051/0004-6361/201527500}, 587, L6

\bibitem[\protect\citeauthoryear{{Villanova}, {Geisler}, {Gratton}  \&
  {Cassisi}}{{Villanova} et~al.}{2014}]{Villanova_2014}
{Villanova} S.,  {Geisler} D.,  {Gratton} R.~G.,   {Cassisi} S.,  2014, \mn@doi
  [\apj] {10.1088/0004-637X/791/2/107}, \href
  {https://ui.adsabs.harvard.edu/abs/2014ApJ...791..107V} {791, 107}

\bibitem[\protect\citeauthoryear{{Weinberg} et~al.,}{{Weinberg}
  et~al.}{2021}]{Weinberg_2021}
{Weinberg} D.~H.,  et~al., 2021, arXiv e-prints, \href
  {https://ui.adsabs.harvard.edu/abs/2021arXiv210808860W} {p. arXiv:2108.08860}

\bibitem[\protect\citeauthoryear{{Wilson} et~al.,}{{Wilson}
  et~al.}{2019}]{Wilson_2019}
{Wilson} J.~C.,  et~al., 2019, \mn@doi [\pasp] {10.1088/1538-3873/ab0075},
  \href {https://ui.adsabs.harvard.edu/abs/2019PASP..131e5001W} {131, 055001}

\bibitem[\protect\citeauthoryear{Yong et~al.,}{Yong et~al.}{2014}]{Yong_2014}
Yong D.,  et~al., 2014, \mn@doi [Monthly Notices of the Royal Astronomical
  Society] {10.1093/mnras/stu118}, 439, 2638–2650

\bibitem[\protect\citeauthoryear{de Boer, Tolstoy, Lemasle, Saha, Olszewski,
  Mateo, Irwin  \& Battaglia}{de~Boer et~al.}{2014}]{deBoer_2014}
de Boer T. J.~L.,  Tolstoy E.,  Lemasle B.,  Saha A.,  Olszewski E.~W.,  Mateo
  M.,  Irwin M.~J.,   Battaglia G.,  2014, \mn@doi [Astronomy & Astrophysics]
  {10.1051/0004-6361/201424119}, 572, A10

\makeatother
\end{thebibliography}

%\begin{thebibliography}{99}
%\end{thebibliography}

%%%%%%%%%%%%%%%%%%%%%%%%%%%%%%%%%%%%%%%%%%%%%%%%%%

%%%%%%%%%%%%%%%%% APPENDICES %%%%%%%%%%%%%%%%%%%%%

\appendix

\section{Some extra material}
\label{appendix1}

\begin{figure*}
\centering
\includegraphics[width=0.74\linewidth]{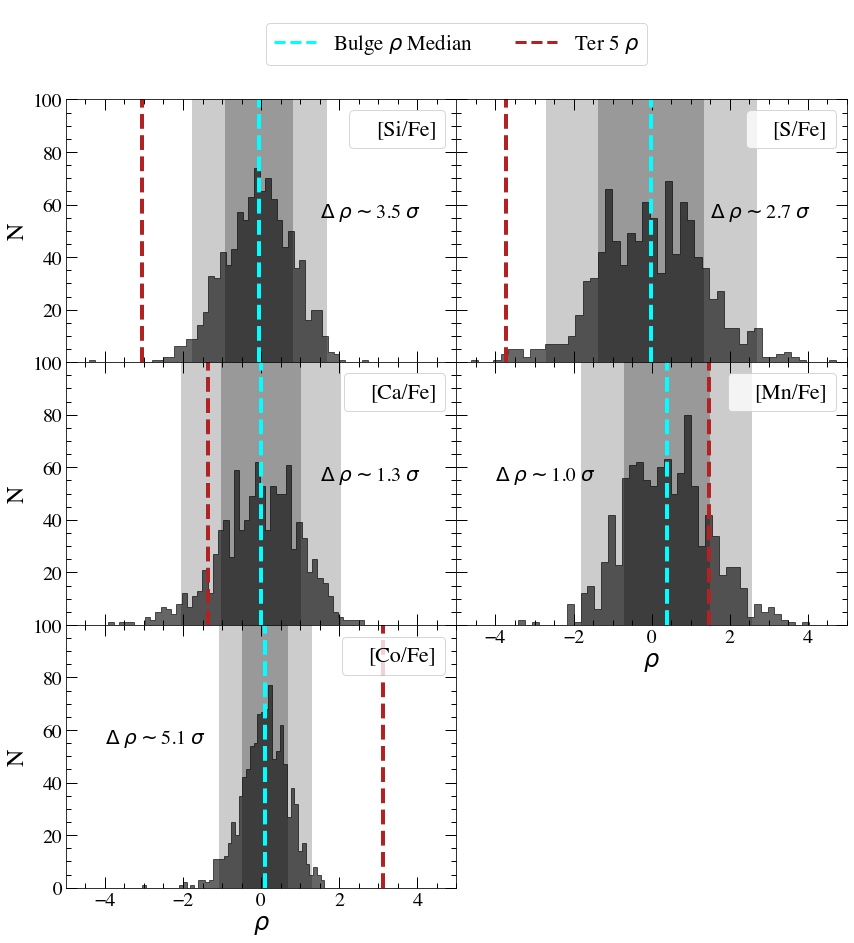}
\caption{Histograms of the $\rho$-distributions of the selected random bulge samples for abundances {\it not} affected by the MP phenomenon, for the strictest sample consisting of 5 candidate Terzan~5 stars. Each panel includes the corresponding median of the randomly sampled bulge $\rho$-distribution (cyan dashed line) -- with light and dark shading indicating its 1$\sigma$ and 2$\sigma$ error, respectively -- and Terzan~5 $\rho$ (red dashed line), along with their separation, $\Delta \rho$.}\label{FigA1}
\end{figure*}

\begin{figure*}
\centering
\includegraphics[width=0.74\linewidth]{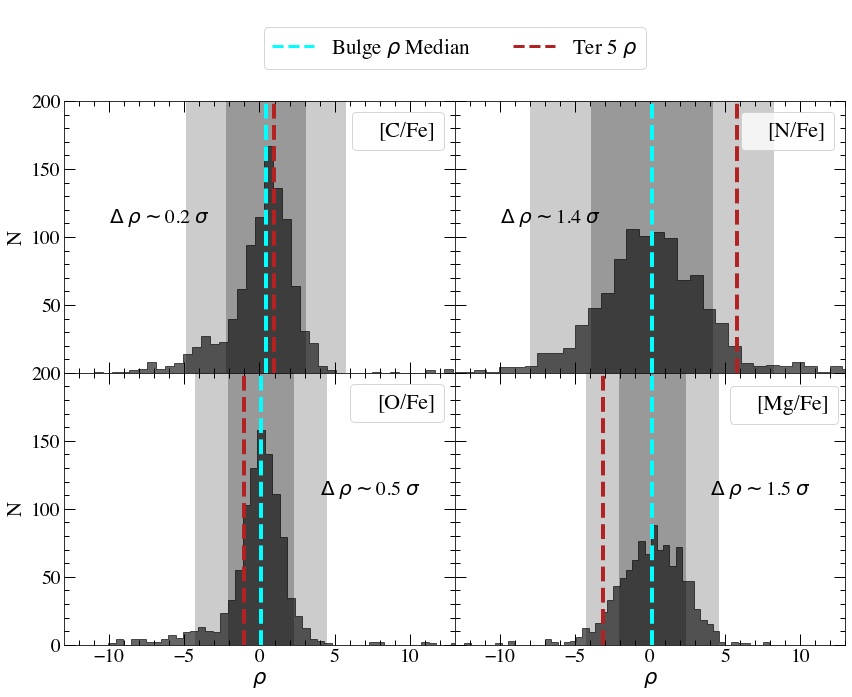}
\caption{Histograms of the $\rho$-distributions of the selected random bulge samples for abundances affected by the MP phenomenon, for the strictest sample consisting of 5 candidate Terzan~5 stars.  The result for Al is omitted for the 5-star sample due to the absence of available Terzan~5 abundances.  Each panel includes the corresponding median of the randomly sampled bulge $\rho$-distribution (cyan dashed line) -- with light and dark shading indicating its 1$\sigma$ and 2$\sigma$ error, respectively -- and Terzan~5 $\rho$ (red dashed line), along with their separation, $\Delta \rho$.}\label{FigA2}
\end{figure*}

%%%%%%%%%%%%%%%%%%%%%%%%%%%%%%%%%%%%%%%%%%%%%%%%%%

% Don't change these lines
\bsp	% typesetting comment
\label{lastpage}
\end{document}